\documentclass[%
reprint,  
superscriptaddress,
amsmath,
amssymb,
aps,
pra,
floatfix,
]{revtex4-2}
\usepackage{graphicx} 
\usepackage{dcolumn} 
\usepackage{bm}
\usepackage{hyperref}
\usepackage{xcolor}
\usepackage{xfrac}
\usepackage{cleveref}
\usepackage{physics}

\usepackage{float}      

\newcommand{\affchem}{Department of Chemistry, Purdue University, West Lafayette, Indiana 47907, USA}
\newcommand{\affphys}{Department of Physics and Astronomy, Purdue University, West Lafayette, Indiana 47907, USA}

\begin{document}

\preprint{APS/123-QED}

\title{Narrow-Line Electric Quadrupole Cooling And Background-Free Imaging Of A Single Cs Atom With Spatially Structured Light}

\author{Karl N. Blodgett}
\altaffiliation{These authors contributed equally to this work.}
\affiliation{\affchem}

\author{Saumitra S. Phatak}
\altaffiliation{These authors contributed equally to this work.}
\affiliation{\affphys}

\author{Meng Raymond Chen}
\affiliation{\affchem}
\author{David Peana}
\affiliation{\affchem}
\author{Claire Pritts}
\affiliation{\affphys}
\author{Jonathan D. Hood}
\email{hoodjd@purdue.edu}
\affiliation{\affchem}
\affiliation{\affphys}
\altaffiliation{$^*$Corresponding author: hoodjd@purdue.edu}

\date{\today}
 
\begin{abstract}
We demonstrate background-free imaging and sideband cooling of a single $^{133}$Cs atom via the narrow‐line 6S$_{1/2}\to5$D$_{5/2}$ electric quadrupole transition in a 1064 nm optical tweezer. The 5D$_{5/2}$ state decays through the 6P$_{3/2}$ state to the ground state, emitting an 852 nm wavelength photon that allows for background-free imaging. By encoding both spin and orbital angular momentum onto the 685 nm excitation light, we achieve background‐free fluorescence histograms with 99.58(3)\% fidelity by positioning the atom at the dark center of a vortex beam. Tuning the tweezer polarization ellipticity realizes a magic trap for the stretched $\lvert F\!=\!4,m_F\!=\!4\rangle\to\lvert F'\!=\!6,m_F'\!=\!6\rangle$ cycling transition. We cool to $5\,\mu$K in a 1.1 mK trap and outline a strategy for ground-state cooling. We compare cooling performance across different sideband regimes, while also exploring how the orbital angular momentum of structured light controls the selection rules for quadrupole transitions.  These results expand the toolbox for high-fidelity quantum control and cooling in alkali-atom tweezer arrays.
\end{abstract}

\maketitle

\section{Introduction}

\begin{figure*}[t!]
    \centering
    \includegraphics[width=\textwidth]{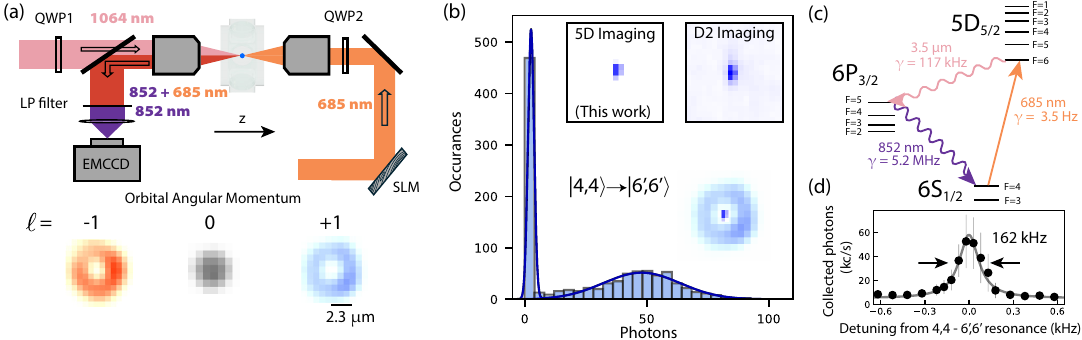}
    \caption{\label{fig1} {Background-free electric quadrupole imaging of $^{133}$Cs with spatially structured light.} (a, top) Optical setup used to address the 6S$_{1/2}$–5D$_{5/2}$ transition. The 1064 nm tweezer and 685 nm excitation light are delivered through high NA objectives situated symmetrically about a UHV glass cell. The ellipticity of the 1064 nm tweezer beam is controlled by a quarter wave plate (QWP1).
    The 685 nm beam reflects off a phase-only spatial light modulator, which imprints a helical phase mask to create vortex beams with OAM of $\ell=-1$, 0, or 1. (a, bottom) Images of the 685 nm beam. A quarter wave plate (QWP2) then adds SAM of $\sigma=\pm1$. This light is focused through the back objective to a 3.2 $\mu$m beam waist, with its center aligned with the trapped atom.
    (b) Background-free imaging histogram of fluorescence from a single Cs atom using 5D light. Photons are scattered by placing the atom at the dark center of the vortex beam. Fitting to a bimodal distribution yields imaging fidelity of 0.9958(3). Inset compares background scattering associated with quadrupole (left)  and standard D2 (right) imaging. (c) Cesium level structure.  
    The 6S$_{1/2} \rightarrow$ 5D$_{5/2}$ electric quadrupole transition (685 nm, $\Gamma \approx$ 117~kHz) leads to cascade decay through 6P$_{3/2}$, emitting an 852 nm photon. This wavelength difference enables background-free imaging. (d) Frequency scan of the $|4,4\rangle \rightarrow |6',6'\rangle$ transition fluorescence, revealing a power-broadened linewidth of 162 kHz.
    }    
\end{figure*}

Ultracold alkali atoms have become the workhorse platform for quantum technologies, enabling practical implementations of atomic clocks ~\cite{Bregazzi2024Acoldatom, wang2023first, gerginov2025accuracy}, quantum sensors ~\cite{Dominik2024Quantum, Ye2024Quantum}, and neutral atom arrays for quantum simulations ~\cite{Saffman2010Quantum, Bernien2017Probing, manovitz2025quantum}. Maintaining ultracold temperatures is essential for the performance and stability of all these applications, but standard laser cooling techniques for alkali atoms rely on the bright D1 and D2 transitions, which face fundamental temperature limits due to their large linewidths ($\sim$5 MHz). Overcoming these limits typically requires additional sub-Doppler schemes such as polarization gradient cooling, Raman sideband cooling, or evaporative cooling~\cite{javanainen1981laser, leibfried2003quantum,  Phatak2024Generalized}. 

In the atomic ion community, narrow-line electric quadrupole (E2) transitions have long been used for ground state preparation~\cite{diedrich1989laser, roos1999quantum, eschner2003laser}. More recently, experimentalists have used narrow optical transitions in alkaline-earth(like) atoms for sideband cooling in magic traps~\cite{saskin2019narrow,norcia2018microscopic,norcia2019seconds,cooper2018alkaline}, and for attractive and repulsive Sisyphus cooling in traps with non-zero differential polarizabilities between the ground and excited states~\cite{covey20192000, cooper2018alkaline, urech2022narrow, ivanov2011laser, taieb1994coolinga, wineland1992sisyphus,graham2023midcircuit, berto2021prospects, Holzl2023Motional, biagioni2025narrowline}. 
In this paper, we demonstrate neutral alkali atom cooling and background-free imaging in Cs using a narrow-line E2 transition. The $6S_{1/2}\rightarrow5D_{5/2}$ transition has a wavelength of 685 nm and decays with a linewidth of 117.6(4)~kHz~\cite{pucher2020lifetime} through the $6P_{3/2}\rightarrow6S_{1/2}$ cascade, emitting an 852 nm photon that can be used for background-free imaging where the excitation light is filtered from the fluorescence~\cite{carr2014improving} (see Fig.~\ref{fig1}(c)).  This narrow line has been proposed for cesium- and rubidium-based optical clocks with reduced size, weight, and power requirements~\cite{sharma2022analysis,duspayev2024optical}.

Electric dipole (E1) transitions are mediated by the electric field through the interaction potential $V_D = d_i E_i$, where $d_i = -e x_i$ couples states of opposite parity, and $x_i$ is the electron position operator along each of the three spatial axes. In contrast, electric quadrupole transitions arise from the gradient of the field via $V_Q = Q_{ij}\, \partial_i E_j$, where the quadrupole moment $Q_{ij} = \frac{e}{2} {x}_i {x}_j$ connects states of the same parity.


Laguerre-Gaussian (LG) modes can drive E2 transitions, which involve a change of two units of angular momentum, as they can simultaneously carry orbital angular momentum (OAM) through their transverse field gradient~\cite{allen1992orbital} and spin angular momentum (SAM) through circular polarization.
In these beams, the phase of the electric field winds around the beam propagation axis, where each photon carries orbital angular momentum of $\ell\hbar$, where $\ell$ is the azimuthal mode number. The OAM of such beams has been used to induce rotational motion in a variety of systems, including nanoparticles~\cite{he1995direct}, macroscopic objects~\cite{padgett2011tweezers,stilgoe2022controlled}, and Bose-Einstein condensates~\cite{andersen2006quantized}. The quantum analogue has been demonstrated by transferring OAM from light to both the valence electron and motional states of an atomic ion via an E2~\cite{verde2023trapped,afanasev2018experimental,stopp2022coherent,quinteiro2017twisted,quinteiro2020paraxial} and an E3~\cite{lange2022excitation} transition.

We report on the use of spatially structured light to achieve background-free imaging with the closed $|F\!=\!4,m_F\!=\!4\rangle\rightarrow|F'\!=\!6,m_F'\!=\!6\rangle$ cycling transition by placing the atom at the dark center of a vortex beam. We investigate cooling across the transverse profile of LG beams with $\ell=-1, 0, +1$. We elucidate the selection rules for quadrupole transitions that depend on both OAM and SAM, and calculate the non-paraxial contributions to the matrix elements. By tuning the polarization ellipticity of the 1064 nm tweezer laser, we achieve magic conditions for the cycling transition and demonstrate single-photon sideband cooling across three regimes: magic trapping where trapping frequencies are matched, attractive Sisyphus cooling where the excited state is more tightly confined, and repulsive Sisyphus cooling where it is more weakly bound. We find optimum cooling in the attractive Sisyphus regime, where we cool the atom to 5~$\mu$K in a 1.1 mK trap with corresponding mean thermal radial and axial motional quantum numbers of $\langle n \rangle_\text{rad}=0.8$ and $\langle n \rangle_\text{ax}=4.3$.

These techniques enable enhanced imaging and cooling capabilities that can improve gate fidelity in neutral atom arrays~\cite{levine2018high}, increased efficiency of molecular assembly~\cite{liu2019molecular,zhang2020forming}, and more compact optical clocks~\cite{sharma2022analysis}. The simplified approach of single-photon sideband cooling with a single beam will also advance atom control and cooling near nanophotonic devices~\cite{tiecke2014nanophotonic,hood2016atomatom, zhou2024trapped}.

\section{Quadrupole Imaging} \label{sec:image}
%

A single cesium atom is imaged using the narrow-line 6S$_{1/2}$ $\rightarrow$ 5D$_{5/2}$ electric quadrupole transition at 685 nm~\cite{tojo2004absorption, chan2016doppler}, shown in Fig.~\ref{fig1}(c).  Weak coupling ($\gamma = 3.5 $Hz) results in a large saturation intensity of $\sim$4 W/cm$^2$.  The 5D$_{5/2}$ state decays with a linewidth of 117.6(4)~kHz~\cite{pucher2020lifetime} through the 6P$_{3/2}$ state, which then emits an 852 nm (D2) photon that is detected by our camera.  The 5D$_{5/2}$ state has inverted hyperfine levels $F'=1$ to $6$.  Due to $F$ selection rules, the $F'=6$ level can only decay to the 6P$_{3/2}$ $F=5$ level, which then decays to the ground state 6S$_{1/2}$ $F=4$, resulting in a closed transition. 

 Fig.~\ref{fig1}(a) depicts our experimental setup. We trap a single Cs atom in a 1064 nm optical tweezer inside an ultra-high vacuum (UHV) glass cell. The 685 nm laser is locked to a temperature-stabilized cavity and is frequency-tuned through two AOMs (see appendix for more details).  
In order to reach the high saturation intensity, the cavity-locked 685 nm light is focused down to a beam waist of 3.2 $\mu$m, where we can generate intensities much larger than $I_\text{sat}$ of approximately 3 kW/cm$^2$ with less than 1 mW optical power. The 1064 nm tweezer and 685 nm excitation light are delivered through NA$=0.55$ objectives situated symmetrically about the UHV glass cell. 

The tweezer polarization ellipticity is controlled by sending the 1064 nm light through a quarter wave plate (QWP1), after which it is focused to a 0.95 $\mu$m beam waist by the front objective.
The  685 nm beam is reflected off of a phase-only spatial light modulator (SLM), which imprints a tunable helical phase mask \(e^{il \phi} \) that results in azimuthal phase winding and OAM of order $\ell$.  When focused through the back objective, the beam adopts a Laguerre-Gaussian modal profile in the focal plane. Throughout this paper, we employ LG$^{\ell}_{0}$ modes with $\ell = -1, 0, 1$, which carry corresponding OAM values of $-1, 0, 1$ with respect to our lab $z$-axis, defined by the tweezer laser wave vector, $\mathbf{k}_\text{tweezer}$. The bottom panel of Fig.~\ref{fig1}(a) displays images taken of the three LG$^{\ell}_{0}$ beam profiles used to drive the electric quadrupole transition. 

The polarization of the 685 nm light is controlled by a quarter waveplate (QWP2), which imparts SAM of either $\sigma=-1$ or $+1$ with respect to the lab $z$-axis. The 685 nm light is focused down through the back objective where, unless otherwise stated, it is centered on the single atom trapped in the tweezer. The fluorescent 852 nm photons are collected through the front objective, reflected off of a dichroic mirror, and imaged onto an EMCCD camera.  Several 852 nm line pass (LP) filters only allow the 852 nm light onto the camera. 

The experiment begins by loading a single atom into a 1.4 mK deep optical tweezer from a D2 line MOT using polarization gradient (PG) cooling. The PG light induces light-assisted collisions to parity-project the number of atoms to either 0 or 1. This same PG light is used to image the atom afterward, confirming the presence of a single atom with a loading rate of 60\%. The tweezer ellipticity is adjusted via QWP1 to conditions where the  $|6',6'\rangle$ state is more trapped than the $|4,4\rangle$ ground state, with polarizability ratio $\alpha_{e}/\alpha_{g}=1.25$ 
\begin{figure*}[tb!]
    \centering
    \includegraphics[width=\textwidth]{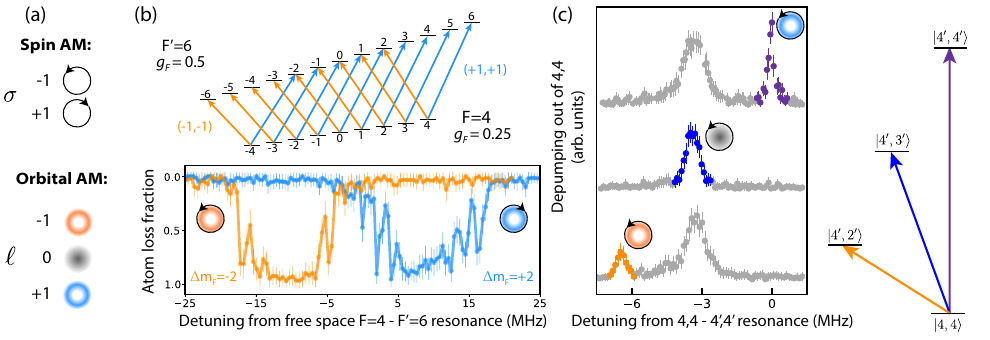}
    \caption{ Quadrupole selection rules. (a) Schematic of the spin and orbital angular momentum that we can controllably encode onto our laser. (b) The top shows the $m_F$ manifold for the $F=4$ and $F'=6$ levels in a magnetic field. The excited state sublevels shift by exactly twice as much as the ground state sublevels. The blue and orange lines draw all of the transitions associated with $\Delta m_F=+2$ and $\Delta m_F=-2$, respectively. The bottom plot shows the experimental atom loss spectra recorded by matching the net angular momentum of the driving laser field to the magnitude of the $|\Delta m_F|=2$ transitions.    (c) Depumping out of the $\left|4,4\right\rangle$ state by matching ($\ell,\sigma$) = (+1, -1) (upper trace), (0, -1) (middle trace), or (-1, -1) (bottom trace). 
    }  
    \label{fig2}
\end{figure*}

To prepare the atom for quadrupole imaging we first optically pump it into the stretched ground state $\left|4, 4\right\rangle$, after which we ramp the tweezer down to a depth of 140 $\mu$K and apply a 6 Gauss magnetic field along the lab $z$-axis. 
The 685 nm laser is configured with ($\ell$, $\sigma$) = (+1, +1), creating light that carries 2 units of total angular momentum with respect to the lab $z$-axis in order to drive the $\left|4,4\right\rangle \rightarrow \left|6',6'\right\rangle$ transition. The frequency of the 685 nm light is tuned to maximize scattering, with a characteristic linewidth shown in Fig.~\ref{fig1}(d). For the histograms and images presented herein, photons are collected for 35 ms at an intensity of 60 $I_\text{sat}$. From our photon collection efficiency of $\sim$3$\%$, we derive a scattering rate of 58 kHz.


Fig.~\ref{fig1}(b) presents a typical imaging histogram collected when the atom is placed into the dark center of the vortex beam. The background free imaging manifests as a bimodal photon count distribution where the background scattering is primarily within the 0-1 photon bin. We emphasize that we have not subtracted any offsets from this histogram. The clear separation of extremely low background scattering and atom fluorescence results in an imaging fidelity of 0.9958(3), which could be pushed even higher by taking more care to filter out residual 685 nm light, or by optimizing cooling parameters during imaging. The inset of Fig.~\ref{fig1}(b) compares images of the atom taken using 5D quadrupole (left) and standard D2 line scattering (right) where the resonant 852 nm imaging light appears as a background. 

While the $F'=6$ state decays only to the $F=4$ ground state, the atom can occasionally off-resonantly scatter from the $F'=5$ excited state, which can decay to the $F=3$ ground state. At saturation intensity, this off-resonant scattering is approximately $10^{-6}$, but since we operate well above saturation, we apply a weak F-repumper addressing the D2, $F=3 \rightarrow F'=4$  transition throughout the imaging sequence to prevent optical pumping into this dark state. Also, to maintain population in the $\left|4,4\right\rangle$ magnetic sublevel during imaging, a weak $\sigma^+$ polarized $m_{F}$-repumper beam resonant with the $F=4 \rightarrow F'=4$ quadrupole transition is applied along the same path as the quadrupole imaging beam. 

The back-to-back imaging survival rate is 0.95(1). For the imaging results presented here, we optimize only for maximum scattering at the center of the vortex beam. Later in this paper, we present a clear strategy for achieving even higher imaging survival rates by optimizing cooling and scattering simultaneously, which occurs for off-axis Gaussian driving light.

\section{Quadrupole selection rules with orbital angular momentum}
\label{sec:quad}
\subsection{Theory}
We now examine the transition matrix elements and selection rules to understand how different $m_F \rightarrow m_F+q$ transitions can be selectively driven. The quadrupole transition has a potential $V_Q = Q_{ij} \, \partial_i E_j$, where the quadrupole $Q_{ij} = \frac{e}{2} x_i x_j$ is in terms of the electron coordinates, and $\partial_i$ is the derivative of the field with respect to the atom position. A quadrupole transition is driven by field gradients rather than the field itself~\cite{carr2014improving,murphree2020quadrupole, bougouffa2020qadrupole, gallagher2025nonparaxial}, which in the classical picture corresponds to a changing field as the atom oscillates, and in the quantum picture corresponds to a field variation over the atom's spatial wavefunction.  It is important to note that the gradient can be either in amplitude or phase, which means that a simple plane wave can drive the transition. 

As derived in the appendix, the transition matrix elements for a quadrupole transition are
\begin{multline}
   \langle F, m | V_Q | F', m' \rangle = \\
    \langle F || \hat{Q}^{(2)} || F' \rangle \, \sum_q (-1)^q \, \langle F, m | F', m'; 2, q \rangle \, (\partial E)^{(2)}_{-q} ,
\end{multline}
where  $ \langle F || \hat{Q}^{(2)} || F' \rangle$ is the quadrupole reduced matrix element, and  $\langle F, m | F', m'; 2, q \rangle $ is the Clebsch-Gordon coefficient with selection rule $m = m' + q$.  The field gradient is a rank 2 spherical tensor that can be written in terms of the rank 1 spherical vectors: 
\begin{align}
    (\partial E )^{(2)}_{\pm 2} &=  \partial_{\pm 1} E_{\pm 1} \label{eq:partial1}\\
    (\partial E )^{(2)}_{\pm 1} &= \frac{1}{\sqrt{2}} ( \partial_0 E_{\pm 1}  + \partial_{\pm 1} E_0 ) \\
    (\partial E )^{(2)}_{0} &= \frac{1}{\sqrt{6}} ( \partial_1 E_{-1} + \partial_{-1} E_{1} + 2 \, \partial_0 E_0 )
\end{align}
The spherical vectors can be expressed in terms of the Cartesian vector as $E_0 = E_z$ and $E_{\pm 1} = \mp  ( E_x \pm i E_y)/\sqrt{2}$, $\partial_0 = \partial_z$, and $\partial_{\pm 1} = \mp (\partial_x \pm i \partial_y)/\sqrt{2}$. 

To understand these selection rules, we first consider a plane wave propagating along the $z$-axis. There is no radial variation of the field and so $\partial_\pm$ is 0. The derivative along the propagation direction $\partial_0 = \partial_z$ is $i k $, which activates three possible quadrupole components: $ (\partial E)^{(2)}_{\pm 1}= \frac{ik }{\sqrt{2}} \,  E_{\pm1}$  and $(\partial E)_0^{(2)} = \frac{2 i k }{\sqrt{6}} \, E_0    $.  These are similar to the dipole case where only the  polarization controls the angular momentum change.

Next we consider structured light, where we have a Laguerre-Gaussian beam LG$_{p=0}^l$ propagating along the $z$-axis. The field near the focus to first order in $r$ and $z$ is:
\begin{equation}
E^{l}_q(r, \theta, z) \approx E_q \left(\frac{r}{w_0}\right)^{|l|} e^{il\theta}\, (1 + ikz)
\end{equation}
For these beams, the directional derivatives selectively couple to the beam's angular momentum: $\partial_{+1} E^l_q$ is non-zero only when $l=+1$, while $\partial_{-1} E^l_q$ is non-zero only when $l=-1$, enabling targeted excitation of specific transitions.

In general, these LG modes drive transitions with $\Delta m_F = q + l$, where $q$ and $l$ represent the photon's spin and orbital angular momentum. While longitudinal phase gradients scale with the wave vector $k$, the radial amplitude gradients scale inversely with the beam waist ($1/\omega_0$), making them weaker by the paraxial factor $s = 1/(\omega_0 k)$.

When a Gaussian beam propagates at an angle relative to all three trap axes, its phase variation projects onto all spatial coordinates. This configuration produces non-zero values for all components of the quadrupole field gradient tensor $(\partial E)^{(2)}_q$, enabling excitation of any allowed quadrupole transition. Experimentally, this represents the most versatile and straightforward approach for driving electric quadrupole transitions, as the strong longitudinal phase gradient ($\sim k$) can couple to any transition when appropriately projected.


\subsection{Experiment}
Fig.~\ref{fig2}(b, upper) shows the level structure diagram for the $F\!=\!4$ and $F'\!=\!6$ levels when the atom is in a magnetic field. Transitions associated with $\Delta m_F=-2$ are drawn as orange arrows and those associated with $\Delta m_F=+2$ are drawn as blue arrows. The data in Fig.~\ref{fig2}(b, lower) shows atom loss spectra for when the atom is prepared in an incoherent mixture of all $m_F$ ground states and strongly driven with 780 $I_{\text{sat}}$ to the $F'=6$ excited state manifold with particular vortex order and polarization. 

For this data, the tweezer was ramped to a low trap depth of 45 $\mu$K and a 6 Gauss B-field was aligned with the 685 nm beam wave vector, $\mathbf{k}_{5D}$. This configuration maximized the alignment of $\mathbf{k}_{5D}$ with the atomic quantization axis, which enabled direct coupling of both spin and orbital angular momentum from the 685 nm photons to the atom. Atom loss is recorded when resonant excitation heats the atom out of the shallow trap. 

The blue spectra was recorded by sending in an LG mode with 
($\ell$, $\sigma$) $= (+1,+1)$
to drive the $\Delta m_F=+2$ transitions.  In contrast, the orange spectra was recorded by sending in ($\ell$, $\sigma$) $=$ (-1, -1) 
light to drive the $\Delta m_F=-2$ transitions.  The frequencies  on the horizontal axis are with respect to the free space $F=4$ $\rightarrow$ $F'=6$ transition.  The two $m_F$-changing transition manifolds exhibit clear symmetric splitting around this central frequency. 

Next, we demonstrate more finely tuned OAM transfer to the atom via transitions within the $\Delta m_F$ manifold of the $F=4 \rightarrow F'=4$ quadrupole transition, as shown in Fig.~\ref{fig2}(c). We use the same spatial arrangement for this measurement as we did for the Zeeman transitions we just considered, i.e., with $\mathbf{k}_{5D}$ aligned along the quantization axis. For each case, we prepare the atom in the $\left|4,4\right\rangle$ ground state and send in 685~nm light carrying $\sigma=-1$ 
and $\ell=$ -1, 0, or +1. The interaction is measured via depumping out of the $\left|4,4\right\rangle$ state.

We focus on three transitions: $\left|4,4\right\rangle \rightarrow \left|4',4'\right\rangle$, $\left|4,4\right\rangle \rightarrow \left|4',3'\right\rangle$, and $\left|4,4\right\rangle \rightarrow \left|4',2'\right\rangle$, shown in purple, blue, and orange, respectively, in Fig.~\ref{fig2}(c). In the bottom spectrum, we send light with ($\ell$, $\sigma$) $=$ (-1,-1) 
character that drives the $\left|4,4\right\rangle \rightarrow \left|4',2'\right\rangle$ transition. In the top spectrum, we send light carrying (1,-1) which drives the $\left|4,4\right\rangle \rightarrow \left|4',4'\right\rangle$ transition. The absence of this $\Delta m_F = 0$ transition in the bottom spectrum, and the absence of the $\Delta m_F = -2$ transition in the top spectrum demonstrates the ability to transfer discrete amounts of OAM from the photon to the atom by only switching the topological charge of the vortex beam from $\ell=1$ to $\ell=-1$. 

The middle spectrum, recorded with ($\ell,\sigma$) = (0, -1) light, shows only the $\ket{4,4} \rightarrow \ket{4',3'}$ transition ($\Delta m_F = -1$), consistent with the angular momentum transfer from the light. The Gaussian beam drives this transition through its longitudinal phase gradient along $z$. Interestingly, this $\Delta m_F = -1$ transition appears prominently in all three spectra, even for vortex beams where it should be suppressed at the beam center. This suggests the atom was not perfectly centered in the vortex beams during these measurements, allowing the strong longitudinal gradient component to drive this transition regardless of beam type.

\begin{figure*}[t]
    \centering
    \includegraphics[width=1.0\linewidth]{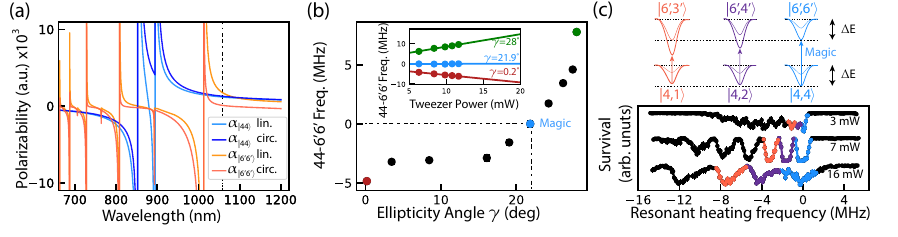}
    \caption{Magic trapping. (a) Polarizability of the $\left|4,4\right\rangle$ and $\left|6',6'\right\rangle$ states in linear and circular tweezer light. Our trapping wavelength of 1064 nm is marked with a dashed line. (b) $\left|4,4\right\rangle$ $\rightarrow$ $\left|6',6'\right\rangle$ resonant scattering frequency shift vs. tweezer polarization ellipticity. The magic condition is found at an ellipticity angle of 21.9$^\circ$. Inset shows the differential trap frequency for magic tweezer, and ellipticity-tuned to either side of magic. (c) Atom loss spectra vs. resonant heating frequency for the $F=4$ $\rightarrow$ $F'=6$ manifold at the magic tweezer ellipticity for three tweezer powers. The ellipticity-tuned magic stretched state transition (blue) frequency remains unchanged with tweezer power, while all other transitions experience a differential light shift. The blue and orange peaks, for example, correspond to the $\left|4,2\right\rangle$ $\rightarrow$ $\left|6',4'\right\rangle$ and $\left|4,1\right\rangle$ $\rightarrow$ $\left|6',3'\right\rangle$ transitions, respectively, which are shown schematically above.
    }
    \label{fig3}
\end{figure*}

\section{Magic polarization} \label{sec:magic}
The ground and excited states of neutral atoms generally have different polarizabilities. For tightly trapped atoms, this difference can result in heating or atom loss, with the effects becoming more significant as the lifetime of the excited state becomes comparable to the period of the trapping frequency. Single-photon resolved sideband cooling requires that the potentials are similar, or magic, so that the two potential wells share a similar set of harmonic oscillator states.  However, a slight mismatch has been shown to improve cooling performance in attractive or repulsive Sisyphus cooling~\cite{covey20192000, cooper2018alkaline, ivanov2011laser}.  

Fig.~\ref{fig3}(b) shows the experimentally measured $\left|4,4\right\rangle$ $\rightarrow$ $\left|6',6'\right\rangle$ transition frequency as a function of the trap laser polarization ellipticity. We find that this frequency crosses zero, i.e., is magic, at a trap polarization ellipticity angle of 22$^\circ$. The inset plots scattering frequency vs. tweezer power at three ellipticity-tuned conditions; $\alpha_{\left|4,4\right\rangle{}}$ \(>\) $\alpha_{\left|6',6'\right\rangle{}}$ (top trace), $\alpha_{\left|4,4\right\rangle{}}$ \(=\) $\alpha_{\left|6',6'\right\rangle{}}$ (middle trace), $\alpha_{\left|4,4\right\rangle{}}$ \(<\) $\alpha_{\left|6',6'\right\rangle{}}$ (bottom trace), where the slopes are proportional to the differential trap depth of the levels.  

While Cs 6S$_{1/2}$ $\rightarrow$ 5D$_{5/2}$ does have a magic wavelength at 803~nm\cite{carr2016doubly, sharma2022analysis}, we use the polarization ellipticity of the tweezer to make the $\ket{4,4}$ and $\ket{6',6'}$ levels magic.
The dynamic polarizabilities for the $\ket{4,4}$ and $\ket{6',6'}$ states are plotted as a function of wavelength in Fig.~\ref{fig3}(a). We plot the polarizabilities for both linear and circular polarized light. 
The polarizabilities are calculated using matrix elements from Ref.~\cite{SafronovaDatabase2022} to find the scalar $\alpha^s(\lambda)$, vector $\alpha^v(\lambda)$, and tensor $\alpha^T(\lambda)$ shifts.  The $m_F$ energies are then calculated by diagonalizing the matrices~\cite{LeKien2013Dynamical}: 
\begin{align}
    V(\lambda)= -\frac{1}{4} |E|^2  \left[  \alpha^s(\lambda) - i \alpha^v(\lambda)  \frac{(\mathbf{\epsilon}^* \times \mathbf{\epsilon}) \cdot \mathbf{F}  }{2F}  +  \right.  \nonumber \\
    \left.   \alpha^T(\lambda)  \frac{ 3  \left( (\mathbf{\epsilon}^* \cdot F)(\mathbf{\epsilon}\cdot F) + (\mathbf{\epsilon} \cdot F)( \mathbf{\epsilon}^*\cdot F) \right) - 2 \mathbf{F}^2   }{2F(2F-1)}   \right] 
\end{align}
Here $\mathbf{\epsilon}$ is the complex polarization of the trapping light.  The vector shift term depends on the ellipticity $-i (\mathbf{\epsilon}^* \times \mathbf{\epsilon})$, which vanishes for linear polarization and is maximized for circular polarization.  The ground state in alkali atoms do not have a tensor shift. 

We plan to use the $\ket{4,4} \rightarrow \ket{6',6'}$ transition for resolved sideband cooling, which requires these states to have equal polarizabilities (magic condition). There are two strategies to achieve this. The conventional approach uses linearly polarized trapping light at a wavelength where the polarizabilities match. However, we adopt an alternative strategy: adjusting the ellipticity of the 1064 nm tweezer polarization to use the vector light shift to equalize the potentials. At 1064 nm, the polarizabilities for the $\ket{4,4}$ state fall between those of the $\ket{6',6'}$ state, guaranteeing that a specific ellipticity exists that makes them magic. Our calculations predict this condition occurs at an ellipticity angle of $15^\circ$.


Our ability to nullify the differential light shift for the stretched state is further illustrated in Fig.~\ref{fig3}(c, lower), where the atom loss spectra is plotted for three tweezer trap depths at the magic condition. We observe a shift in all transitions except for the magic-tuned stretched state (highlighted in blue). Fig.~\ref{fig3}(c, upper) shows a color-coded potential energy cartoon of the corresponding transitions. The $\left|4,4\right\rangle$ and $\left|6',6'\right\rangle$ tweezer potentials shift by the same amount, whereas all other transitions have a nonzero differential light shift. 

The discrepancy between the calculated magic angle of 15$^\circ$ and experimentally measured angle of 22$^\circ$ could be due to a combination of possible factors. First, polarization at the atom might differ from that measured by our polarimeter~\cite{schaefer2007measuring}, possibly due to birefringence of our objective. Second, there may be inaccuracies in the matrix elements used to calculate the 5D$_{5/2}$ state polarizability, as evidenced by the historical discrepancy between its measured and calculated lifetimes~\cite{diberardino1998lifetime,pucher2020lifetime, Hoeling1996Lifetime}. 

Additionally, the paraxial factor of our tightly-focused tweezer beam is $1/(\omega_0 k) = 0.18$. Using the Bluestein method~\cite{hu2020efficient}, we calculate the non-paraxial focal region electric field vectors (shown in appendix). For our 50~$\mu$K atoms with harmonic oscillator lengths of 250~nm (axial) and 70~nm (radial), longitudinal field components reach a maximum of only $\sim$1\% at spatial extrema. Therefore, non-paraxial effects cannot explain the discrepancy between measured and calculated magic polarizability.


In the above analysis we have ignored the polarizability contribution of the 685 nm light, as its intensity is three orders of magnitude lower than that of the tweezer light at the atom. We also assume that the intermediary 6P$_{3/2}$ state, with short lifetime of $\sim$30 ns relative to typical trapping frequency periods of $\sim$10 $\mu$s, can be ignored.

\begin{figure}[t]
    \centering
    \includegraphics[width=1\linewidth]{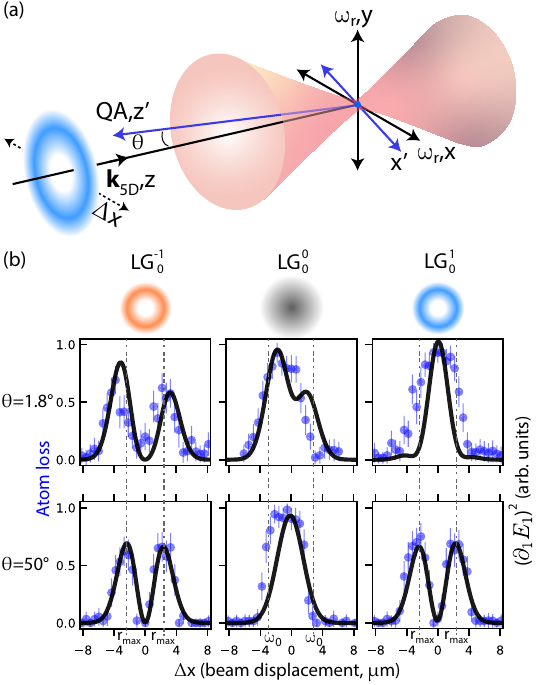}
    \caption{Position-resolved interaction. (a) Schematic of the experimental geometry of the 685 nm excitation light ($\mathbf{k}_{5D}$, blue vortex beam) propogating along $z$, at an angle $\theta$ with respect to the atomic quantization axis, which can lie in the xz plane. (b) Measured $n\rightarrow n$ interaction strength as a function of beam displacement, $\Delta x$ for three beam modes; LG$^{\text{$-1$}}_{\text{$0$}}$ (left), LG$^{\text{$0$}}_{\text{$0$}}$ (middle), and LG$^{\text{$1$}}_{\text{$0$}}$ (right) for a small $\theta$ (top row) and large $\theta$ (bottom row). The calculated squared value of the interaction term, $\partial_1 E_1$, is plotted as the black trace. Dashed vertical lines denote the bright ring radius and beam waist for the vortex and Gaussian beams, respectively. }
    \label{fig4}
\end{figure}


\section{Beam Position Effects on Quadrupole Coupling} 
\label{se:spatially}
We now characterize the strength of the $\left|4,4\right\rangle$ $\rightarrow$ $\left|6',6'\right\rangle$ transition versus the beam position. So far we have examined atoms at the center of the 685 nm beam with the quantization axis along the $z$ axis, parallel to $\mathbf{k}_{5D}$.

In Fig.~\ref{fig4}, we scan the 685 nm beam across the atom. In all cases, we drive the $\left|4,4\right\rangle$ to $\left|6',6'\right\rangle$ transition with $\sigma=+1$ while measuring atom loss via resonant heating. The transition is driven by the field gradient component $\partial_1 E_1   = (\partial_x + i \partial_y)(E_x + i E_y )/2$ (see Eq.~\ref{eq:partial1}). The three columns represent different LG beams. In the first row of (b), the quantization axis is nearly aligned with $\mathbf{k}_{5D}$, while in the second row, the quantization axis is rotated by $50^{\circ}$ relative to $\mathbf{k}_{5D}$.  In all plots, the black line is the theoretical prediction.

Fig.~\ref{fig4}(a) shows a schematic for the setup described here.
The small angle regime is characterized by a very low trap depth (45 $\mu$K) and a 6 G magnetic field along the z axis, $B_z$, which is nearly collinear with the 685 nm wave vector, $\mathbf{k}_{5D}$. In this case, the quantization axis (QA) is predominantly defined by $B_z$. We define $\theta$ as the angle between $\mathbf{k}_{5D}$ and the QA (Fig.~\ref{fig4}(a)). 

The top row of Fig.~\ref{fig4}(b) shows the small angle atom loss scans for the $\sigma=+1$ and  $\ell= -1, 0, 1 $
, from left to right.
For each case, we calculate the $\partial_1E_1$ matrix element with respect to the QA using the full vector electric field, and plot its scaled squared magnitude as the solid black trace. We find good qualitative agreement between experiment and theory at an angle $\theta$=$1.8^\circ$.

The $\ell=+1$ data shows that the quadrupole transition is driven strongly by the OAM at its dark center, which arises purely from azimuthal phase winding.  For a vortex beam at the center, $(\partial_1 E_1) $ is non-zero.

In contrast, the Gaussian beam drives the transition most strongly by its transverse amplitude gradients away from the beam center. The non-zero signal at the Gaussian center is due to the longitudinal phase derivative projecting amplitude onto the rotated QA $x'$ and $y'$ axes, where it has a large $\partial_1$ term to drive the transition. 

The $\ell=-1$ 
beam causes no interaction at its dark center due to the mismatch of its phase winding with the transition, for which $(\partial_1 E_1) = 0$. 
The interaction turns on symmetrically at a distance $r$ that extends beyond the beams bright ring radius, $r_{max}$, where the transverse amplitude gradients of the Gaussian envelope project onto Cartesian derivatives $\partial_x$, $\partial_y$, which in turn project onto $\partial_1$. We note that these amplitude gradients are present in the spatially extended $\ell=+1$ 
beam as well, but their effects are overwhelmed by the much stronger phase gradient at the beam's dark center. 

Next, we examine the case where the 5D beam is at a $50^{\circ}$ angle from the QA. For this scenario, we use a trap depth of 1 mK and a zeroed magnetic field. The tweezer polarization is set to $15^{\circ}$ to make the $\ket{4,4}$ and $\ket{6',6'}$ states nearly magic. The vector shift from the tweezer ellipticity produces an artificial magnetic field along the $z$-axis. When combined with the large tensor shift, the overall quantization axis lies in the $xz$-plane.

The bottom row of Fig.~\ref{fig4}(b) shows the large angle atom loss data. We again find good qualitative agreement between experiment and theory at an angle $\theta = 50^{\circ}$. The transition strength in this case follows the beam intensity profile for each OAM state. As the beam is tilted relative to the QA, the rapidly changing phase gradient along the propagation direction projects onto the radial axes. With a paraxial factor of $1/(\omega_0 k) = 0.03$, this phase propagation dominates over the radial amplitude gradients.



For our beam parameters, once there is deviation from the angle $\theta$ beyond $\sim5^\circ$ the longitudinal phase projection onto the rotated QA components becomes the main contributor to the transition (see appendix for detailed theory and spatially-dependent matrix element plots at several values of $\theta$). This highlights the relatively small parameter space in which radial gradients can be experimentally observed --- a regime that can be expanded by increasing the beam's paraxial factor via tighter focusing.

\section{Quadrupole Narrow-line Cooling: Theory}\label{sec:narrow}
Having established magic conditions for the cycling transition, we now turn to investigating single-photon sideband cooling. This technique has two key requirements: the resolved regime and the Lamb-Dicke regime. In the resolved regime, the excited state linewidth must be smaller than the trap frequency to resolve the motional sidebands. The Lamb-Dicke regime requires that the Lamb-Dicke parameter $\eta = k x_0 < 1$ (where $x_0$ is the zero-point oscillation length), meaning that the recoil energy from spontaneous emission is less than the trap frequency.

Alkali atoms traditionally lack suitable narrow-line transitions, necessitating two-photon cooling schemes like polarization gradient, gray molasses, EIT cooling, or Raman sideband cooling for sub-Doppler temperatures~\cite{Phatak2024Generalized}. However, when an excited state's linewidth approaches the trap frequency ($\sim$100 kHz in optical tweezers), single-photon sideband cooling becomes possible, enabling transitions from $|g, n\rangle$ to $|e, n-1\rangle$.

While alkaline-earth atoms have achieved near-ground-state cooling with their narrow intercombination lines, this has not been previously demonstrated in alkali atoms with resolved sideband cooling. Single-photon schemes offer significant experimental advantages: they require only one beam, making them more robust against alignment issues compared to multi-beam approaches that rely on precise polarization gradients or beam overlap.

Although cesium does possess a dipole-allowed transition between $6S_{1/2}$ and $7P_{3/2}$, its linewidth ($\sim$1 MHz) is an order of magnitude larger than typical trap frequencies in optical tweezers, making it unsuitable for sideband cooling~\cite{Toh2019Measurement}. This contrasts with lithium, where the $2S\rightarrow3P$ transition (700 kHz) has been used for cooling free-space MOTs~\cite{Duarte2011All}, but in that case gray molasses techniques yield similar cooling results~\cite{grier2013lambda,salomon2014gray}.

The electric quadrupole interaction arises from the coupling between the atom's quadrupole moment and the gradient of the electric field. For a single atom, this interaction can be written as:
\begin{equation}
V_Q = Q_{ij}\,\partial_i E_j(\hat{\mathbf{r}}),
\end{equation}
where $Q_{ij}$ is the electric quadrupole tensor operator, $\partial_i = \frac{\partial}{\partial x_i}$ denotes the spatial derivative, and $\hat{\mathbf{r}}$ is the position operator of the atom.

To understand sideband cooling, we need to incorporate the quantized motion of the atom in the trap. We represent the atom's state using both internal degrees of freedom, the hyperfine states $|F, m_F\rangle$,  and external degrees of freedom, the harmonic oscillator states $|\mathbf{n}\rangle$ with $\mathbf{n} = (n_x, n_y, n_z)$. 

Each spatial coordinate operator can be expressed in terms of harmonic oscillator raising and lowering operators:
\begin{equation}
\hat{x}_j = x_{0,j}(\hat{a}_j + \hat{a}_j^\dagger), \quad x_{0,j} = \sqrt{\frac{\hbar}{2m\omega_j}},
\end{equation}
where $\omega_j$ is the trap frequency along the $j$-axis and $x_{0,j}$ is the characteristic length scale of the ground state wavefunction.

For neutral atoms trapped in light, the ground and excited states generally experience different trap potentials. The excited state motional wavefunctions $|\mathbf{m}_e\rangle$ can be related to the ground state basis $|\mathbf{n}\rangle$ through a squeezing transformation:
\begin{equation}
|\mathbf{m}_e\rangle = \hat{S}(\bm{\xi})|\mathbf{n}\rangle, \quad \hat{S}(\bm{\xi}) = \exp\left[\sum_j \frac{\xi_j}{2}(\hat{a}_j^2 - \hat{a}_j^{\dagger 2})\right],
\end{equation}
where $\xi_j = \frac{1}{2}\ln\left(\frac{\omega_{e,j}}{\omega_{g,j}}\right)$ characterizes the ratio between excited and ground state trap frequencies. This squeezing parameter arises from the different polarizabilities of the ground and excited states.

With these foundations established, we can express the transition matrix element as:
\begin{align}
&\langle F,m_F;\mathbf{n}|V_Q|F',m_F';\mathbf{m}_e'\rangle =   \nonumber \\
&\langle F\Vert Q\Vert F'\rangle \sum_q (-1)^q\langle F,m_F;2,q|F',m_F'\rangle\langle\mathbf{n}|(\partial E)^{(2)}_{-q}(\hat{\mathbf{r}})|\mathbf{m}_e'\rangle
\end{align}

For small displacements ($\eta \ll 1$) and small differences in trap frequencies ($\xi_j \ll 1$), we can expand the motional part of this matrix element to first order to get the following three terms:
\begin{align}
&\langle\mathbf{n}|(\partial E)^{(2)}_q(\hat{\mathbf{r}})\hat{S}(\mathbf{\xi})|\mathbf{n}'\rangle \approx \delta_{\bm n, \bm n'} (\partial E)^{(2)}_q(0)   \nonumber \\
&+ \sum_i \, \frac{\xi_i}{2} \, \langle\mathbf{n}| (\hat{a}_i^2 - \hat{a}_i^{\dag 2}) |\mathbf{n}'\rangle \,\,  (\partial E)^{(2)}_q(0) \\
&\;+\sum_j \eta_{j}\, \langle\mathbf{n}| (\hat{a}_j + \hat{a}_j^{\dag} )|\mathbf{n}'\rangle
\Bigl. \,\,\frac{\partial}{\partial x_j}(\partial E)^{(2)}_q\Bigr|_{\mathbf{r}=0}
\end{align}
Here we ignore the fourth term ($\xi_j$$\eta$) as it is second order in the expansion.

This result reveals three distinct driving mechanisms: 

1. The first term drives carrier transitions ($\mathbf{n} \rightarrow \mathbf{n}$) with no change in motional state.

2. The second term, proportional to the squeezing parameter $\xi_j$, couples states that differ by two motional quanta ($\Delta n_j = \pm 2$). This mechanism becomes important when the ground and excited state trap frequencies differ significantly.

3. The third term drives first-order sideband transitions ($\Delta n_j = \pm 1$) and depends on the gradient of the electric field derivative. These transitions scale with the Lamb-Dicke parameter $\eta_{j} = k x_{0,j}$.

In our experiment, we exploit these mechanisms to achieve efficient cooling by operating in regimes where one or more of these pathways is enhanced. We note that the quantum picture of attractive and repulsive sisyphus cooling is just sideband cooling with slightly relaxed motional state selection rules, where $\Delta n_j = 0,\pm 1,\pm2$ transitions are all allowed. 

\section{Quadrupole Narrow-line Cooling: Experimental Results}\label{sec:cool-exp}
\begin{figure*}[bt!]
    \centering
    \includegraphics[width=1.0\linewidth]{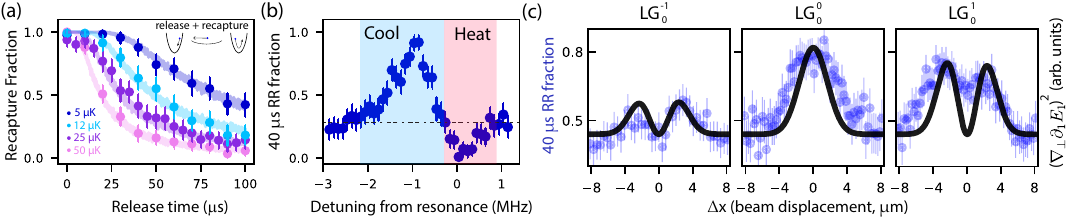}
    \caption{Cooling a single Cs atom using the quadrupole transition. (a) Release and recapture curves for attractive sideband cooling (dark blue), magic-tweezer tuned unresolved sideband cooling (light blue), repulsive sideband cooling (purple), and no cooling (magenta). (b) 40 $\mu$s release and recapture cooling signal as a function of laser detuning in the attractive sideband regime. Heating occurs at blue, and cooling occurs at red detuning. (c) Position-dependent cooling vs. beam position. The calculated squared value of the dominant cooling term, $\nabla_{\perp} \partial_1E_1$, is plotted as the black trace. }
    \label{fig5}
\end{figure*}
In this section, we investigate single-photon sideband cooling of a trapped cesium atom using the electric quadrupole transition under three distinct regimes: magic, attractive, and repulsive sideband cooling. By tuning the tweezer polarization ellipticity, we control the relative depth of the ground and excited state potentials, which allows us to explore different cooling mechanisms with the same experimental apparatus.

Fig.~\ref{fig5}(a) presents the release-recapture curves that demonstrate the efficacy of each cooling regime compared to an uncooled atom. For this data, the atom trap is diabatically switched off for some release time, after which the trap light is turned back on. The atom recapture probability is a statistical measure of the atom radial temperature in the tweezer.

The attractive sideband regime (dark blue) achieves the lowest temperature of 5 $\mu$K, with corresponding mean radial and axial thermal motional quantum numbers of $\langle n \rangle_\text{rad}=0.8$ and $\langle n \rangle_\text{ax}=4.3$. Without cooling (magenta), the atoms remain at their initial temperature of approximately 50 $\mu$K.  The magic trap condition (light blue) reaches 12 $\mu$K with $\langle n \rangle_\text{rad}=2.4$ and $\langle n \rangle_\text{ax}=11$, while the repulsive regime (purple) results in 25 $\mu$K with $\langle n \rangle_\text{rad}=5.6$ and $\langle n \rangle_\text{ax}=23$. 

Fig.~\ref{fig5}(b) demonstrates the frequency dependence of the cooling process in the attractive regime, revealing that red-detuned light produces cooling while blue-detuned light causes heating. Fig.~\ref{fig5}(c) shows how cooling efficiency varies with beam position for different beam modes, revealing the spatial dependence of the cooling interaction.

Our cooling experiments were conducted in a regime approaching resolved sidebands ($\Gamma/\omega_{\text{radial}} = 1.4$) using a 1.1 mK deep trap with zeroed magnetic field and trap frequencies of $\omega_r =$ 86 kHz and $\omega_{\text{ax}} =$ 22 kHz. For this configuration, we extract a value for $\theta$ of 25$^\circ$. For all measurements, we applied 50 ms pulses of Gaussian cooling light at 60 $I_\text{sat}$ while optimizing the frequency to maximize the release-recapture signal.


\vspace{-4mm}
\subsection{Magic Condition}
We first examine cooling when the ground and excited state potentials have equal depths, the so-called magic condition where the polarizabilities ratio $\alpha_e/\alpha_g = 1$. At this condition, the trap frequencies for ground and excited states match, eliminating differential light shifts. Optimal cooling occurs at a detuning of $\delta = -6.9 \, \Gamma$ from the atom-loss resonance, rather than the $-\Gamma$ detuning typical for resolved sideband cooling. This larger detuning is necessary because our system operates outside the fully resolved sideband regime, and the higher cooling power enables more effective energy extraction.

The magic condition provides the simplest form of sideband cooling, which relies on momentum transfer from photons to atoms where the only allowed cooling transition is $\Delta n = -1$. Since we are in the unresolved regime, we anticipate the cooling to be mediocre at best, and for cooling to drastically improve in the resolved sideband regime. Indeed, our calculations indicate that a Gaussian beam with wave vector along a trap axis has large radial and axial cooling matrix elements at an angle $\theta=25^\circ$ (see appendix). 


\vspace{-4mm}
\subsection{Attractive Regime}
By tuning the tweezer polarization ellipticity such that the excited state potential becomes deeper than the ground state by a factor $\alpha_{e}/\alpha_{g}=1.25$, we enter the attractive cooling regime. In the classical picture, cooling requires excitation of the atom from the ground state potential minimum to the excited state potential minimum. The atom must have sufficient time to move away from the trap minima before decay, thereby transferring potential energy to the emitted photon. These criteria typically require operation in the resolved regime, where the trapping frequency in the excited state exceeds the linewidth ($\omega_e\geq\Gamma$).

Optimal cooling in this regime occurs at a detuning $\delta=-12$ $\Gamma$ relative to the atom-loss frequency. The superior cooling performance in this regime (5 $\mu$K versus 12 $\mu$K in the magic case) can be attributed to an additional cooling mechanism beyond pure sideband cooling: the squeezing cooling mechanism. Since the potentials have different curvatures, the excited state wavefunction is squeezed compared to the ground state, enabling transitions that change the vibrational quantum number by two ($\Delta n = \pm 2$). This process is  independent of the projection of photon momentum and therefore works effectively even with our small paraxial factor, and outside of the resolved regime. It remains an open question how cold one can radially and axially cool an atom with attractive sideband cooling, in both the unresolved and resolved regimes. 

\vspace{-4mm}
\subsection{Repulsive Regime}
We also investigated the repulsive regime by setting $\alpha_{e}/\alpha_{g}=0.75$, making the excited state potential shallower than the ground state. Classically, cooling in this regime works by exciting atoms near the classical turning points of the ground state potential. If the atom has time to oscillate before emitting a photon, it can emit more energy than was absorbed during excitation. This mechanism creates an energy threshold (the ``sideband cap") below which atoms are cooled and above which they are expelled from the trap.

In this regime, optimal cooling occurred at a detuning of $\delta=-5.1\Gamma$ from the atom-loss resonance. While this regime achieved the highest final temperature (25 $\mu$K) among the three, it provides valuable insight into the range of cooling mechanisms accessible with quadrupole transitions.

\vspace{-4mm}
\subsection{Spatially Dependent Cooling}

To understand how beam structure affects cooling efficiency, we investigated the spatial dependence of cooling for different Laguerre-Gaussian modes. Using parameters optimized for attractive sideband cooling in a 0.43 mK trap with a 6 Gauss magnetic field along the $z$-axis, we scanned the LG$^{\ell}_{0}$ beams across the atom while measuring the 40 $\mu$s release-recapture survival as an indicator of cooling efficiency. As shown in Fig.~\ref{fig5}(c), for all three beam modes, cooling efficiency correlates directly with beam intensity at our operating angle of $\theta = 25^\circ$ between the beam propagation direction and the quantization axis. 

Since the release-recapture signal is sensitive to the radial temperature of the atom in the tweezer, we restrict our analysis to the radial motional coupling terms, i.e., we compare the positon-dependent cooling signal with the radial gradient of the $ \partial_1E_1$ matrix element, $\nabla_{\perp} \, \partial_1E_1$. The scaled squared value of this is plotted as the black trace in Fig.~\ref{fig5}(c). We note that the squeezing matrix element, which goes with $ \partial_1E_1$, also contributes to cooling in the attractive sideband regime, but for our experimental conditions $\eta$$\nabla_{\perp} \partial_1E_1$ $\sim$ 30$\times$     $\xi \, \partial_1E_1$. Therefore, in Fig.~\ref{fig5}(c) we only consider the dominant cooling term. We note that both cooling mechanisms exhibit nearly identical spatial profiles at such large operating $\theta$ values (see appendix).  

For small angles between the beam and quantization axis ($\theta < 5^\circ$), the calculated cooling profiles exhibit more complex structure, reflecting the second-order spatial derivatives of the electric field. However, achieving such small angles experimentally would require impractically large magnetic fields to overcome the trap-induced quantization axis. Detailed calculations of the spatially-dependent transition matrix elements at various angles are provided in the Appendix.


In our experiment, $\mathbf{k}_{5D}$ is aligned with the axial axis of the tweezer, and makes an angle $\theta$ of 25$^\circ$ with respect to the QA. In this configuration, we find optimal cooling at the center of the Gaussian beam.  Even with minimal momentum projection onto the tweezer radial axes with our largely paraxial cooling beam, our calculations suggest large cooling matrix elements along all three tweezer motional axes for both sideband and squeezed cooling terms. 

The more widely used and experimentally robust configuration is to send the Gaussian cooling beam at an angle with respect to all three trap axes, similar to the configuration used by the ion-trapping community for resolved sideband cooling with electric quadrupole transitions. The large phase gradient projects onto all three trap dimensions, thereby turning on the $\partial_1$ component, which when combined with linear polarized light can drive the stretched $\lvert4,4\rangle\!\to\!\lvert6',6'\rangle$ transition via the $\partial_1 E_1$ term. We plan to use this configuration in a follow up study to do resolved sideband cooling with the E2 transition.

\section{Conclusion} \label{sec:conclusion}
We have demonstrated narrow-line and background-free quadrupole cooling and imaging of a single alkali atom. Tuning the polarization ellipticity of the tweezer enabled precise control of differential light shifts, allowing us to realize a magic trap for the stretched $\lvert4,4\rangle\!\to\!\lvert6',6'\rangle$ cycling transition. In the attractive sideband regime, we cooled the atom to $5\,\mu\mathrm{K}$ in a 1.1 mK deep trap, corresponding to mean thermal motional occupation numbers of $\langle n\rangle_\text{rad}=0.8$ and $\langle n\rangle_\text{ax}=4.3$, despite operating outside the resolved sideband regime. 

We also demonstrated orbital angular momentum transfer by driving the quadrupole transition manifold with spatially structured light, enabling selective excitation of transitions through tailored combinations of OAM and SAM. We measured the dependence of the $\lvert4,4\rangle\!\to\!\lvert6',6'\rangle$ interaction strength and cooling efficiency on the transverse spatial profile of three Laguerre-Gaussian beams, as well as their projection onto the quantization axis, and found good agreement between experiment and theory using non-paraxial vector electric fields.

While our Laguerre-Gaussian beam study revealed fundamental aspects of angular momentum transfer, future implementations could simply use a Gaussian beam with a propagation vector aligned to project onto all three trap axes, enabling simultaneous coupling for scattering, magic sideband cooling, and squeezed sideband cooling. Entering the fully resolved sideband regime by increasing trap depth well beyond 1 mK could enable both radial and axial resolution, potentially allowing near-unit ground state occupation---albeit requiring careful management of the resulting tensor light shift and $m_F$ repumping. 

These results establish a compact, single-beam approach for sideband cooling and background-free imaging in alkali atoms. They point toward simplified protocols for imaging and cooling in tweezer arrays, with potential applications in optical clocks~\cite{sharma2022analysis, carr2016doubly} and other platforms where scattered light or constrained optical access make traditional techniques impractical. Our approach is particularly well suited for integration with nanophotonic quantum devices~\cite{tiecke2014nanophotonic, hood2016atomatom, zhou2024trapped}, where background-free imaging and robust single-beam cooling provide key advantages.

\vspace{3mm}
\textbf{\emph{Acknowledgments:}}
We thank Mark Saffman and Arghavan Safavi‑Naini for valuable discussions on the 5D transition and Chen‑Lung Hung for his careful reading of the manuscript and helpful comments.
We are grateful to the Jonathan Amy Facility for Chemical Instrumentation for its contributions to the instrumentation, and to Lee Liu and Jonah Quirk for guidance on laser locking.
We thank Alexander Urech for fruitful discussions on cooling with a narrow‑linewidth transition.
This work was supported by NSF CAREER Award No. 0543784.

\clearpage
\newpage
\section{Appendix}

\subsection{Cs 5D Laser Lock}
The $6S_{1/2}\rightarrow5D_{5/2}$  signal was initially found with fluorescence from a Cs vapor cell heated to 80 degrees Celsius. We found the transition by scanning our laser around the frequency reported in Sebastian Pucher's thesis of 437.599208 THz~\cite{pucher2018spectroscopy}. Light was collected through a 1 inch diameter triplet collimator, and fiber coupled to a single photon counting module. On our wavemeter, we measure the center of the doppler-broadened transition (600 MHz FWHM) to be 437.59825 THz.  The $|F\!=\!4\rangle\rightarrow|F'\!=\!6\rangle$ transition is approximately 90 MHz red detuned to the center of the doppler-broadened transition.  For reference, we measure the D2 $|F\!=\!3\rangle\rightarrow|F'\!=\!2/3\rangle$ crossover transition frequency to be 351.73075 THz.

We lock our 685 nm diode laser to a nearby fringe of a temperature-stabilized, ultra-low expansion high-finesse cavity ($\mathcal{F}$ = 8,000, FSR=2 GHz). To address the $|F\!=\!4\rangle\rightarrow|F'\!=\!6\rangle$ transition, we use two double pass acousto-optic modulators in series with center frequencies of 80 MHz and 200 MHz. This configuration allows us to scan the laser frequency by hundreds of MHz while maintaining large intensities at the atom.

\subsection{Theory for Quadrupole Transitions}
\subsubsection{Dipole transition }
Before looking at the quadrupole transition, we first review the dipole transition. The dipole interaction potential is given by:
\begin{equation}
    V_D =  \mathbf{d} \cdot \mathbf{E}(\mathbf{r}) ,
\end{equation}
where the dipole operator is proportional to the electron position, ${d}_i = -e x_i$, and $\mathbf{r}$ is the position of the atom.

We calculate the transition matrix elements for a dipole transition, such as the 6S to 6P transition in cesium. The scalar product of the dipole potential in spherical vector form is
\begin{equation}
    V_D =  \sum_q (-1)^q\,    d_{q} \, E_{-q}(\mathbf{r}) \,,
\end{equation}
where $q = 0, \pm 1$ corresponds to the rank 1 spherical vector components. For transitions between atomic hyperfine states $\lvert F, m_F\rangle$ and $\lvert F', m_F'\rangle$, the matrix element is
\begin{multline}
    \langle F, m_F | V_D | F', m_F' \rangle  \\
    = \sum_q (-1)^q \, 
      \langle F, m_F | {d}_q | F', m_F' \rangle  \, E_{-q}(\mathbf{r}) \,.
\end{multline}

Using the Wigner-Eckart theorem, we can separate the the transition matrix element into a $m_F$ independent reduced matrix element and $m_F$ dependent Clebsch-Gordon coefficient:
\begin{equation}
    \langle F, m_F | {d}_q | F', m_F' \rangle 
    = \langle F || {d} || F' \rangle\,
      \langle F, m_F | F', m_F';\, 1, q \rangle,
\end{equation}
where $\langle F || {d} || F' \rangle$ is the reduced matrix element and $\langle F, m_F | F', m_F';\, 1, q \rangle$ is the Clebsch-Gordan coefficient.

These Clebsch-Gordan coefficients give rise to the selection rules for dipole transitions:
\begin{equation}
    \Delta F = 0, \pm 1 
    \quad \text{and} \quad 
    \Delta m_F = q = 0, \pm 1.
\end{equation}
In the dipole case, only the polarization or spin angular momentum (SAM) couples to the system.

\subsubsection{Quadrupole}
For transitions between states of the same parity (e.g., S to D), the dipole transition is forbidden. The next allowed transition arises from the quadrupole moment of the atom. The quadrupole interaction is given by \cite{wooley2020power,murphree2020quadrupole} 
\begin{equation}\label{eq:VQ}
    V_Q = \sum_{ij} Q_{ij} \,\partial_i E_j(\mathbf{r}),
\end{equation}
where $\partial_i = \tfrac{d}{dx_i}$ is the derivative with respect to the atoms position. The quadrupole tensor is a symmetric second-order tensor:
\begin{equation}
    Q_{ij} = \frac{e}{2}\,{x}_i \,{x}_j,
\end{equation}
where $x_i$ is the electron coordinate.

Eq.~\ref{eq:VQ} is the scalar product of two rank-2 Cartesian tensors. We can treat the electric field gradient as a Cartesian rank-2 tensor:
\begin{equation}
    (\partial E)_{ij} = \partial_i E_j.
\end{equation}
This matrix is traceless because the electric field is divergence-free when no free charges are present: $ \partial_i E_i = 0$.

Next, we calculate the quadrupole transition matrix elements between the states $\lvert F,m\rangle$ and $\lvert F',m'\rangle$:
\begin{equation}
    \langle F,m | V_Q | F',m'\rangle
 =  \langle F,m |\sum_{ij} Q_{ij} (\partial E)_{ij}  | F',m'\rangle
\end{equation}
The first step is to decompose the rank-2 Cartesian tensors into the irreducible components of a spherical tensor operator. For example, any rank-2 Cartesian tensor operator $Q_{ij}$ can be decomposed into irreducible spherical tensors of ranks $k=0,1,2$, where:
\begin{itemize}
    \item Rank 0: $Q^{(0)}$ corresponds to the trace of the tensor, which is a scalar.
    \item Rank 1: $Q^{(1)}_{q}$ is related to the antisymmetric part of the tensor, which has three components.
    \item Rank 2: $Q^{(2)}_{q}$ is related to the remaining traceless, symmetric matrix $Q_{ij}$, which has five components.
\end{itemize}
After decomposing both the field gradient and the quadrupole into their irreducible spherical tensors, the scalar product in $V_Q$ can be expressed as the sum of scalar spherical products for each rank:
\begin{equation}
    V_Q = \sum_{ij} Q_{ij}\,(\partial E)_{ij}
        = \sum_k (-1)^k\,Q^{(k)}\cdot(\partial E)^{(k)}.
\end{equation}

In our case, the only term that is non-zero is the rank $k=2$ term. 
This is because of two reasons.  First, The rank-0 term vanishes because $(\partial E)_{ij}$ is traceless and $(\partial E)^{(0)}=0$.  Second, $Q_{ij}$ is symmetric and so its rank 1 component is zero, $Q^{(1)}_q = 0$.  

Consequently, we are left with only the rank-2 scalar product:
\begin{equation}
    V_Q = Q^{(2)} \cdot (\partial E)^{(2)}
         = \sum_q (-1)^q \,Q^{(2)}_q\,(\partial E)^{(2)}_{-q}.
\end{equation}

Now we can return to the transition matrix element. For a transition between hyperfine states $\lvert F, m\rangle$ and $\lvert F',m'\rangle$, the matrix elements is:
\begin{multline}
    \langle F, m | V_Q | F', m' \rangle \;=\;\\
    \sum_q (-1)^q\,\langle F,m | Q^{(2)}_q | F',m' \rangle\,\,(\partial E)^{(2)}_{-q} \;=\;\\
    \langle F || {Q} || F' \rangle
    \sum_q (-1)^q\,\langle F,m | F',m';\,2,q \rangle  \,\,
    (\partial E)^{(2)}_{-q},
\end{multline}
The last line uses the Wigner-Eckart theorem:
\begin{equation}
    \langle F,m | {Q}^{(2)}_q | F',m' \rangle
    \;=\;
    \langle F ||{Q}^{(2)} || F' \rangle
    \,\,\langle F,m | F',m';\,2,q \rangle
\end{equation}

The Clebsch-Gordan coefficient leads to the following selection rules for quadrupole transitions:
\begin{itemize}
    \item $\Delta F = 0,\,\pm 1,\,\pm 2 \quad (\text{excluding }F=0\leftrightarrow F=0),$
    \item $\Delta m = 0,\,\pm 1,\,\pm 2.$
\end{itemize}

The reduced matrix element can be further decomposed using the relationship between hyperfine and fine-structure states:
\begin{multline}
    \langle F || Q^{(2)} || F' \rangle
    \;=\;
    \langle J || Q^{(2)} || J' \rangle\,
    (-1)^{F' + J + 2 + I}\\
    \times \sqrt{(2F'+1)\,(2J+1)}\,
    \begin{Bmatrix}
       J & J' & 2\\
       F' & F & I
    \end{Bmatrix},
\end{multline}
where $\{\cdots\}$ denotes a Wigner 6-j symbol.

The last step is to find an expression for the rank 2 spherical tensor components for $(\partial E)^{(2)}_q$.  This tensor is related to the traceless symmetric components of the matrix $\partial_i E_j + \partial_j E_i$. We now derive an expression for $(\partial E)^{(2)}_{q}$ in terms of the Cartesian tensors $\partial_i E_j$. We can express the rank 2 spherical tensor in terms of the spherical tensor multiplication of the rank 1 tensors $\partial^{(1)}$ and $E^{(1)}$ as
\begin{multline}\label{eq:partialE}
    (\partial E)^{(2)}_q
    = \{\partial^{(1)} \otimes E^{(1)}\}^{(2)}_q \\
    = \sum_{q_1,\,q_2}
    \partial_{q_1}\,E_{q_2}
    \,\langle 1,q_1;\,1,q_2 \mid 2,q\rangle.
\end{multline}  
The spherical components of the fields and their derivatives are related to the Cartesian components by
\begin{align}
    E_0 &= E_z,\\
    E_{\pm1} &= \mp\frac{1}{\sqrt{2}}\,(E_x \pm i\,E_y),
\end{align}
and similarly for the derivatives.

By inserting the relevant Clebsch-Gordan coefficients, we can expand Eq.~\eqref{eq:partialE} into
\begin{align}
    (\partial E)^{(2)}_{\pm 2}
      &= \partial_{\pm 1}\,E_{\pm 1},\\
    (\partial E)^{(2)}_{\pm 1}
      &= \frac{1}{\sqrt{2}}\bigl(
         \partial_0\,E_{\pm 1} \;+\; \partial_{\pm 1}\,E_0
         \bigr),\\
    (\partial E)^{(2)}_0
      &= \frac{1}{\sqrt{6}}\bigl(
         \partial_1\,E_{-1} \;+\; \partial_{-1}\,E_1
         \;+\; 2\,\partial_0\,E_0
         \bigr).
\end{align}

In Cartesian coordinates, these become
\begin{align}
    (\partial E)^{(2)}_0 
      &= \sqrt{\frac{3}{2}}\left[
         \partial_z E_z
         \;-\;\frac{1}{3}\bigl(\partial_x E_x + \partial_y E_y + \partial_z E_z\bigr)
         \right],\\
    (\partial E)^{(2)}_{\pm1}
      &= \mp\,\frac{1}{2}\Bigl[
         (\partial_x E_z + \partial_z E_x)\,\pm\, i\,(\partial_y E_z + \partial_z E_y)
         \Bigr],\\
    (\partial E)^{(2)}_{\pm2}
      &= \frac{1}{2}\Bigl[
         \partial_x E_x \;-\; \partial_y E_y
         \;\pm\; i\,\bigl(\partial_x E_y + \partial_y E_x\bigr)
         \Bigr].
\end{align}

\subsubsection{Plane wave along the $z$-axis}
For a plane wave propagating along the $z$ quantization axis,
\[
{E}_q(\mathbf{r}) = {\epsilon}_q\, e^{ikz}.
\]
The electric field depends only on the coordinate $z$. Consequently, every spatial derivative except $\partial_0 = \partial_z$ vanishes.  The $z$ derivative gives $\partial_z e^{ikz} = i k e^{ikz}$.  The resulting quadrupole fields are 
\begin{align}
(\partial E)^{(2)}_{\pm1} &= \frac{1}{\sqrt{2}} \bigl( i k E_{\pm1} \bigr),\\
(\partial E)^{(2)}_{0} &= \frac{2}{\sqrt{6}} \bigl( i k E_{0} \bigr).
\end{align}
Thus, the selection rules reduce to the familiar dipole rules, with the change of $m_F$ fixed by the light polarization $E_q$.

\subsubsection{Laguerre--Gaussian beam along $z$-axis}
The Laguerre-Gaussian (LG) modes in cylindrical coordinates are:
\begin{align}
E_{p}^{\,l}(r,\theta,z) &= 
E_0\;
\frac{w_0}{w(z)}
\left(\frac{\sqrt{2}r}{w(z)}\right)^{|l|}
L_{p}^{|l|}\!\!\left(\frac{2r^{2}}{w^{2}(z)}\right) \nonumber \\
&\times \exp\!\left[-\frac{r^{2}}{w^{2}(z)}\right] 
\exp\!\left[ikz + ik\frac{r^{2}}{2R(z)}\right. \nonumber \\
&\left. - i(2p+|l|+1)\arctan\left(\frac{z}{z_{R}}\right) + il\theta\right]
\end{align}
where $w(z) = w_{0}\sqrt{1+(z/z_{R})^{2}}$, $R(z) = z(1+(z_{R}/z)^{2})$, and $z_{R} = \pi w_{0}^{2}/\lambda$. Here $l$ is the OAM index, $p$ is the radial index, $w_{0}$ is the beam waist, and $L_{p}^{|l|}$ is the associated Laguerre polynomial.

For our experiments with $p=0$ modes, $L_0^{|l|}(x) = 1$. Near the focus ($r,z \ll w_0, z_R$), the first order field simplifies to:
\begin{equation}
E^{l}_0(r, \theta, z) \approx E_0 \left(\frac{\sqrt{2} r}{w_0}\right)^{|l|} e^{il\theta} (1 + ikz + i (1+|l|) \frac{z}{z_R} )
\end{equation}
In the paraxial regime, $k \gg 1/z_R$ and the Guoy phase in last term can be ignored.
\begin{equation}
E^{l}_0(r, \theta, z) \approx E_0 \left(\frac{\sqrt{2} (x + i y)}{w_0}\right)^{|l|}  (1 + ikz)
\end{equation}

The $l=0$ case is a Gaussian and has selection rules $\Delta m_F = q $, where $q$ is the polarization. For $l=1$, the field is zero at the center, and therefore, all the  $\partial_0$ terms vanish.  However, the radial $\partial_{(\pm 1)}$ term is non-zero for the $l=\pm 1$ terms.  In general, it can drive a transition with $\Delta m_F = l + q $.   Both the polarization (SAM) and orbital angular momentum (OAM) contribute to the total angular momentum change. 


\subsection{Narrow-line cooling}
In the previous section on transition matrix elements, we omitted the effects of photon recoil on the atom's position. We now incorporate the atom's quantized motion to understand sideband cooling mechanisms for both dipole and quadrupole transitions.

We model both ground and excited states as three-dimensional harmonic oscillator states $|\mathbf{n}\rangle$, where $\mathbf{n} = (n_x, n_y, n_z)$ represents the motional quantum numbers along each axis. The position operator for each axis can be expressed in terms of the raising and lowering operators and the characteristic length scale:
\[
\hat{x}_j = x_{0,j}(\hat{a}_j+\hat{a}_j^\dagger), \quad
x_{0,j}=\sqrt{\frac{\hbar}{2m\omega_j}},
\]
where $\omega_j$ is the trap frequency along axis $j$.

In general, the excited state experiences a different trapping potential from the ground state. We denote the excited state harmonic oscillator states as $|\mathbf{m}_e\rangle$, which can be related to ground state basis via a squeezing transformation. For a single dimension, this transformation is given by:
\[
\hat{S}(\xi) = \exp\left[\frac{\xi}{2}(\hat{a}^2-\hat{a}^{\dagger 2})\right], \quad \xi = \frac{1}{2}\ln\left(\frac{\omega_e}{\omega_g}\right),
\]
where $\xi$ characterizes the ratio of excited to ground state trap frequencies. For three dimensions, we define a vector squeezing parameter $\mathbf{\xi}$ corresponding to each axis. For potentials with similar trapping frequencies, the squeezing operator can be expanded as 
\begin{align}
    \hat{S}(\xi) \approx 1 + \frac{\xi}{2} (\hat{a}^2 - \hat{a}^{\dagger 2}),
\end{align}
which couples states $n$ to $n \pm 2$.

The complete basis states now include both internal and motional degrees of freedom: $|F, m_F; \mathbf{n}\rangle$ for the ground state and $|F', m_F'; \mathbf{m}_e\rangle = \hat{S}(\mathbf{\xi})|F', m_F'; \mathbf{n}\rangle$ for the excited state. In the following sections, we analyze sideband cooling mechanisms for both dipole and quadrupole transitions within this framework.

\subsubsection{Dipole Sideband Cooling}
To incorporate the motional coupling, we upgrade the atom position dependence in the electric field to a position operator $\hat{\mathbf{r}}$. The 3D harmonic oscillator states $|\mathbf{n}\rangle$ with $\mathbf{n} = (n_x, n_y, n_z)$ are included with the hyperfine states to get $|F, m_F; \mathbf{n} \rangle$. For details see Ref.~\cite{Phatak2024Generalized}.

The transition matrix element separates into electronic and motional parts: 
\begin{align}
&\langle F,m_F;\mathbf{n} | V_D(\mathbf{r}) | F',m_F';\mathbf{m}_e' \rangle
      \nonumber\\
&\quad=   \sum_q (-1)^q   \langle F, m_F  |d_q | F', m_F'\rangle  \,\, \langle \mathbf{n} | E_{-q}(\hat{\mathbf{r}})   | \mathbf{m}_e' \rangle
\end{align}

The electronic matrix element is the same as in the previous section. The last term gives the motional coupling between the ground $|\mathbf{n} \rangle$ and excited $|\mathbf{m}_e \rangle$ harmonic oscillator states, where we have allowed for generally different trapping frequencies in ground and excited states.

Expanding the field at the trap center to first order in each axis gives:
\begin{align}
E_q(\hat{\mathbf{r}})
&\approx
E_q(0)+\sum_{j=x,y,z} \hat{x}_j\Bigl.\frac{\partial E_q(\mathbf{r})}{\partial x_j}\Bigr|_{\mathbf{r}=0}
\end{align}

The motional coupling between two different harmonic states is:
\begin{align}
&\langle\mathbf{n}|E_q(\hat{\mathbf{r}})|\mathbf{m}_e'\rangle = \langle\mathbf{n}|E_q(\hat{\mathbf{r}})\hat{S}(\mathbf{\xi})|\mathbf{n}'\rangle 
\end{align}

For the case where $\xi_i \ll 1 $ and $\eta \ll 1 $, 
then we can expand $S(\xi)$ and $\eta$ to get 
\begin{align}
&\langle\mathbf{n}|E_q(\hat{\mathbf{r}})\hat{S}(\mathbf{\xi})|\mathbf{n}'\rangle
\approx \delta_{\bm n, \bm n'} E_q(0)   \nonumber \\
&+ \sum_i \, \frac{\xi_i}{2} \, \langle\mathbf{n}| (\hat{a}_i^2 - \hat{a}_i^{\dag 2}) |\mathbf{n}'\rangle \,\,  E_q(0) \\
&\;+\sum_j \eta_j\, \langle\mathbf{n}| (\hat{a}_j + \hat{a}_j^{\dag} )|\mathbf{n}'\rangle
\Bigl. \,\,\frac{\partial E_q}{\partial x_j}\Bigr|_{\mathbf{r}=0}
\end{align}
The first term drives the carrier transition ($\mathbf{n} \rightarrow \mathbf{n}$) through the electric field gradient and is proportional to the zeroth-order term in the Lamb-Dicke expansion and squeezing. The second term drives $n \rightarrow n \pm 2$ transitions through the squeezing parameter $\xi$, which accounts for different trap frequencies in ground and excited states. The third term drives the first-order sidebands ($n \rightarrow n \pm 1$) and scales with the Lamb-Dicke parameter $\eta_{j}$. Finally, the fourth term combines both effects, coupling through both the squeezing parameter and the field gradient, which can be ignored as $\xi$$\eta  \ll$ 1.

\subsubsection{Quadrupole Sideband Cooling}
For an electric–quadrupole transition the interaction is
\begin{equation}
V_Q \;=\; Q_{ij}\,\partial_i E_j(\hat{\mathbf{r}}).
\end{equation}
The matrix element of $V_Q$ factorizes into electronic and motional parts:
\begin{align}
&\langle F,m;\bm n | V_Q | F',m';\bm m_e' \rangle                       \nonumber\\
&\;=\;
\langle F || Q ||F' \rangle
\sum_q (-1)^q
      \langle F,m | F',m', 2,q \rangle                              \nonumber\\
&\quad\;\times
\bigl\langle \bm n \bigl|
     (\partial E)^{(2)}_{-q}(\hat{\mathbf{r}})
 \bigr| \bm m_e' \bigr\rangle .
\label{eq:QME}
\end{align}

Next, we take a Taylor series expansion of the quadrupole field gradient tensor around the trap center $\mathbf{r} = 0$:
\begin{align}
(\partial E)^{(2)}_{q}(\hat{\mathbf{r}}) &\approx (\partial E)^{(2)}_{q}(0) + \sum_{j} \hat{x}_j \left.\frac{\partial}{\partial x_j}(\partial E)^{(2)}_{q}\right|_{\mathbf{r}=0} 
\end{align}
Inserting this between the ground and excited motional states gives 
\begin{align}
&\langle\mathbf{n}|(\partial E)^{(2)}_q(\hat{\mathbf{r}})|\mathbf{m}_e'\rangle = \langle\mathbf{n}|(\partial E)^{(2)}_q(\hat{\mathbf{r}})\hat{S}(\mathbf{\xi})|\mathbf{n}'\rangle 
\end{align}

For the case where $\xi_i \ll 1 $, we can expand $\hat{S}(\mathbf{\xi})$ to get 
\begin{align}
&\langle\mathbf{n}|(\partial E)^{(2)}_q(\hat{\mathbf{r}})\hat{S}(\mathbf{\xi})|\mathbf{n}'\rangle \approx \delta_{\bm n, \bm n'} (\partial E)^{(2)}_q(0)   \nonumber \\
&+ \sum_i \, \frac{\xi_i}{2} \, \langle\mathbf{n}| (\hat{a}_i^2 - \hat{a}_i^{\dag 2}) |\mathbf{n}'\rangle \,\,  (\partial E)^{(2)}_q(0) \\
&\;+\sum_j \eta_{j}\, \langle\mathbf{n}| (\hat{a}_j + \hat{a}_j^{\dag} )|\mathbf{n}'\rangle
\Bigl. \,\,\frac{\partial}{\partial x_j}(\partial E)^{(2)}_q\Bigr|_{\mathbf{r}=0}
\end{align}

The first term drives the carrier transition ($\mathbf{n} \rightarrow \mathbf{n}$) through the quadrupole field gradient and is proportional to the zeroth-order term in the Lamb-Dicke expansion. The second term drives $n \rightarrow n \pm 2$ transitions through the squeezing parameter $\xi$, which accounts for different trap frequencies in ground and excited states. The third term drives the first-order sidebands ($n \rightarrow n \pm 1$) and scales with the quadrupole Lamb-Dicke parameter $\eta_{j}$. The fourth term being second order is ignored similar to the dipole case. These different mechanisms provide multiple pathways for sideband cooling in the quadrupole transition.





\subsection{Quadrupole Interaction Plots}
In this section we present radial cross section plots of several components of the electric quadrupole interaction, where we used the Bluestein method~\cite{hu2020efficient} to compute the vector electric field within a few micrometers of our 685 nm laser beam’s focal region. Large multiplicative factors correspond to small components while small factors correspond to large components.

\begin{figure}[H]
    \centering
    \includegraphics[width=1\linewidth]{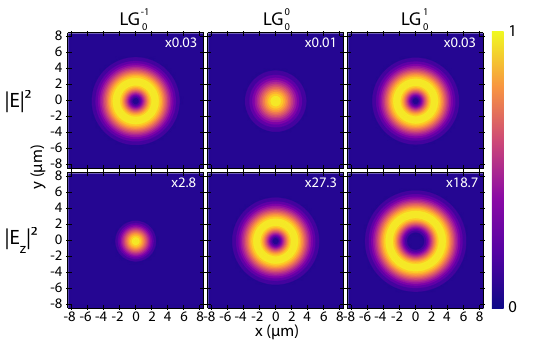}
    \caption{Total and longitudinal intensity profiles for the three Laguerre-Gaussian beams used in this paper. Each beam is encoded with $\sigma$=+1 SAM with respect to our lab $z$-axis. We use our measured beam waist of 3.2 $\mu$m and wavelength of 685 nm. The relatively large longitudinal component at the beam center occurs for the case when OAM and SAM are mismatched, i.e., for $\sigma=-\ell=1$ ~\cite{quinteiro2017twisted}. }
    \label{figA1}
\end{figure}

\begin{figure*}[!ht]
    \centering
    \includegraphics[width=1\linewidth]{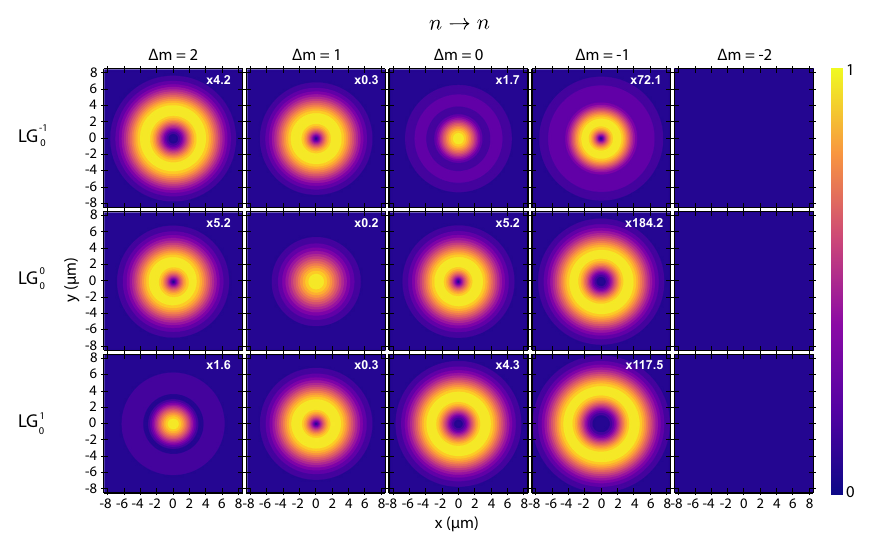}
    \caption{The n$\rightarrow$n carrier transition strength for the $\Delta F=2$ electric quadrupole transition. Each $\Delta m_F$ component is plotted for each of the three beam modes using the electric field in the focal plane with the beam propogating along the $z$-axis. For these plots, the beam is encoded with $\sigma$=+1, and the angle between the 685 nm $\mathbf{k}$ and the QA is zero.    
    }
    \label{figA2}
\end{figure*}

\begin{figure}[H]
    \centering
    \includegraphics[width=1\linewidth]{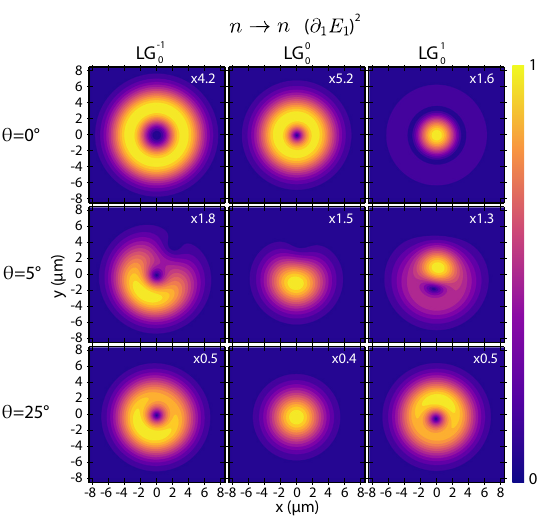}
    \caption{The n$\rightarrow$n carrier transition strength for the $\Delta m_F=+2$ component of the electric quadrupole transition, $(\partial_1E_1)^2$. The interaction profile is plotted for several values of $\theta$, which defines the angle between the 685 nm $\mathbf{k}$ and the QA. In each case, the beam is encoded with $\sigma$=+1, and the 685 nm beam propogates along the $z$-axis. }
    \label{figA3}
\end{figure}

\begin{figure}[H]
    \centering
    \includegraphics[width=1\linewidth]{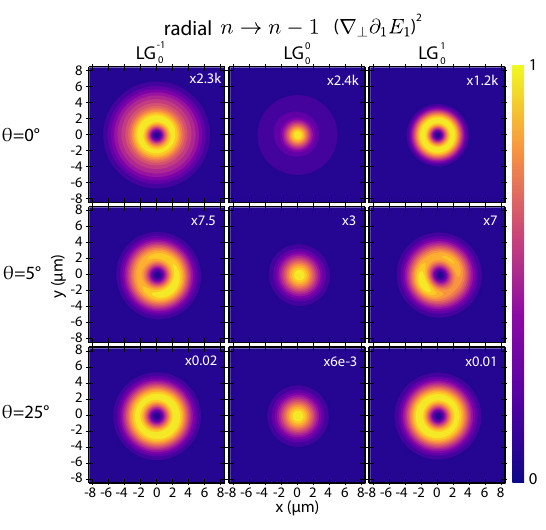}
    \caption{The n$\rightarrow$n-1 radial sideband transition strength for the $\Delta m_F=+2$ component of the electric quadrupole transition, $(\nabla_{\perp}\partial_1E_1)^2$. The interaction profile is plotted for several values of $\theta$, which defines the angle between the 685 nm $\mathbf{k}$ and the QA. In each case, the beam is encoded with $\sigma$=+1, and the 685 nm beam propogates along the $z$-axis. }
    \label{figA4}
\end{figure}

\begin{figure}[H]
    \centering
    \includegraphics[width=1\linewidth]{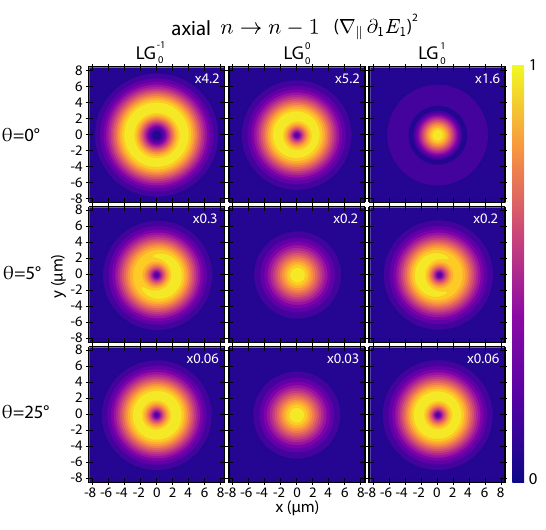}
    \caption{The n$\rightarrow$n-1 axial sideband transition strength for the $\Delta m_F=+2$ component of the electric quadrupole transition, $(\nabla_{\parallel}\partial_1E_1)^2$. The interaction profile is plotted for several values of $\theta$, which defines the angle between the 685 nm $\mathbf{k}$ and the QA. In each case, the beam is encoded with $\sigma$=+1, and the 685 nm beam propagates along the $z$-axis. }
    \label{figA5}
\end{figure}

\subsection{Tweezer Laser Non-paraxial Analysis}

In this section we present transverse and longitudinal electric field components for our 1064 nm gaussian tweezer laser at the beam focus, with a beam waist of 0.95 $\mu$m. We used the Bluestein method~\cite{hu2020efficient}  to compute the vector electric field within a few micrometers of the optical tweezer’s focal region. Large multiplicative factors correspond to small components while small factors correspond to large components.

\begin{figure*}[htbp]
    \centering
    \includegraphics[width=1\linewidth]{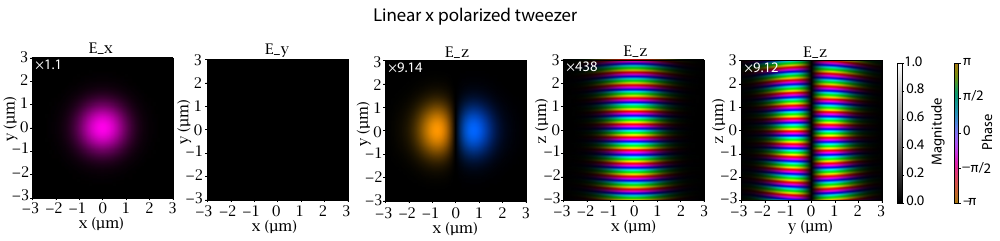}
    \caption{Transverse and longitudinal electric field components at the beam focus for a linearly polarized tweezer laser. }
    \label{figA6}
\end{figure*}

\begin{figure*}[htbp]
    \centering
    \includegraphics[width=1\linewidth]{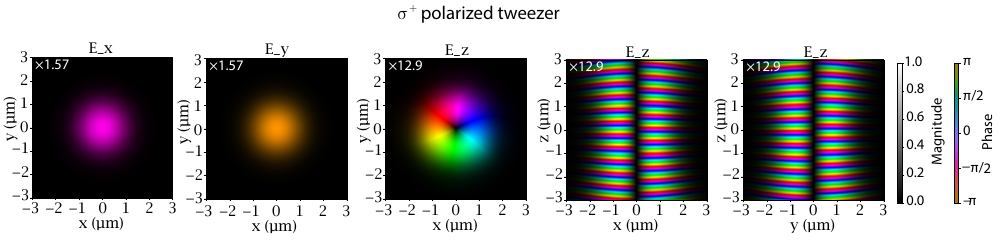}
    \caption{Transverse and longitudinal electric field components at the beam focus for a circularly polarized tweezer laser with $\sigma$=+1. }
    \label{figA7}
\end{figure*}

\begin{figure*}[htbp]
    \centering
    \includegraphics[width=1\linewidth]{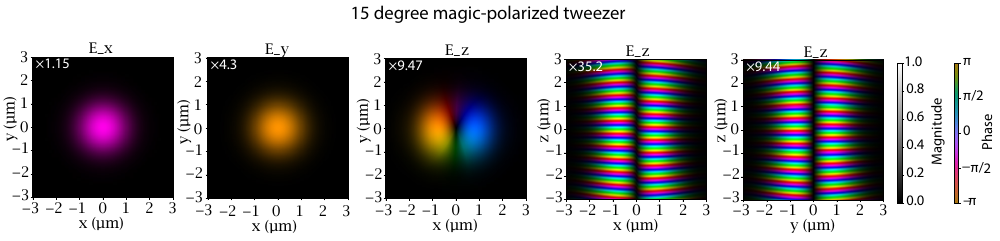}
    \caption{Transverse and longitudinal electric field components at the beam focus for a tweezer laser with polarization ellipticity angle = 15$^\circ$, which is calculated to be magic for the $\left|4,4\right\rangle$ $\rightarrow$ $\left|6',6'\right\rangle$ transition. }
    \label{figA8}
\end{figure*}

\clearpage
\bibliographystyle{apsrev4-2}
\bibliography{references_master.bib}

\providecommand{\noopsort}[1]{}\providecommand{\singleletter}[1]{#1}%
\begin{thebibliography}{66}%
\makeatletter
\providecommand \@ifxundefined [1]{%
 \@ifx{#1\undefined}
}%
\providecommand \@ifnum [1]{%
 \ifnum #1\expandafter \@firstoftwo
 \else \expandafter \@secondoftwo
 \fi
}%
\providecommand \@ifx [1]{%
 \ifx #1\expandafter \@firstoftwo
 \else \expandafter \@secondoftwo
 \fi
}%
\providecommand \natexlab [1]{#1}%
\providecommand \enquote  [1]{``#1''}%
\providecommand \bibnamefont  [1]{#1}%
\providecommand \bibfnamefont [1]{#1}%
\providecommand \citenamefont [1]{#1}%
\providecommand \href@noop [0]{\@secondoftwo}%
\providecommand \href [0]{\begingroup \@sanitize@url \@href}%
\providecommand \@href[1]{\@@startlink{#1}\@@href}%
\providecommand \@@href[1]{\endgroup#1\@@endlink}%
\providecommand \@sanitize@url [0]{\catcode `\\12\catcode `\$12\catcode `\&12\catcode `\#12\catcode `\^12\catcode `\_12\catcode `\%12\relax}%
\providecommand \@@startlink[1]{}%
\providecommand \@@endlink[0]{}%
\providecommand \url  [0]{\begingroup\@sanitize@url \@url }%
\providecommand \@url [1]{\endgroup\@href {#1}{\urlprefix }}%
\providecommand \urlprefix  [0]{URL }%
\providecommand \Eprint [0]{\href }%
\providecommand \doibase [0]{https://doi.org/}%
\providecommand \selectlanguage [0]{\@gobble}%
\providecommand \bibinfo  [0]{\@secondoftwo}%
\providecommand \bibfield  [0]{\@secondoftwo}%
\providecommand \translation [1]{[#1]}%
\providecommand \BibitemOpen [0]{}%
\providecommand \bibitemStop [0]{}%
\providecommand \bibitemNoStop [0]{.\EOS\space}%
\providecommand \EOS [0]{\spacefactor3000\relax}%
\providecommand \BibitemShut  [1]{\csname bibitem#1\endcsname}%
\let\auto@bib@innerbib\@empty
\bibitem [{\citenamefont {Bregazzi}\ \emph {et~al.}(2024)\citenamefont {Bregazzi}, \citenamefont {Batori}, \citenamefont {Lewis}, \citenamefont {Affolderbach}, \citenamefont {Mileti}, \citenamefont {Riis},\ and\ \citenamefont {Griffin}}]{Bregazzi2024Acoldatom}%
  \BibitemOpen
  \bibfield  {author} {\bibinfo {author} {\bibfnamefont {A.}~\bibnamefont {Bregazzi}}, \bibinfo {author} {\bibfnamefont {E.}~\bibnamefont {Batori}}, \bibinfo {author} {\bibfnamefont {B.}~\bibnamefont {Lewis}}, \bibinfo {author} {\bibfnamefont {C.}~\bibnamefont {Affolderbach}}, \bibinfo {author} {\bibfnamefont {G.}~\bibnamefont {Mileti}}, \bibinfo {author} {\bibfnamefont {E.}~\bibnamefont {Riis}},\ and\ \bibinfo {author} {\bibfnamefont {P.~F.}\ \bibnamefont {Griffin}},\ }\href {https://doi.org/10.1038/s41598-024-51418-8} {\bibfield  {journal} {\bibinfo  {journal} {Sci. Rep.}\ }\textbf {\bibinfo {volume} {14}},\ \bibinfo {pages} {931} (\bibinfo {year} {2024})}\BibitemShut {NoStop}%
\bibitem [{\citenamefont {Wang}\ \emph {et~al.}(2023)\citenamefont {Wang}, \citenamefont {Ruan}, \citenamefont {Liu}, \citenamefont {Guan}, \citenamefont {Shi}, \citenamefont {Yang}, \citenamefont {Bai}, \citenamefont {Zhang}, \citenamefont {Fan}, \citenamefont {Wu}, \citenamefont {Zhao},\ and\ \citenamefont {Zhang}}]{wang2023first}%
  \BibitemOpen
  \bibfield  {author} {\bibinfo {author} {\bibfnamefont {X.}~\bibnamefont {Wang}}, \bibinfo {author} {\bibfnamefont {J.}~\bibnamefont {Ruan}}, \bibinfo {author} {\bibfnamefont {D.}~\bibnamefont {Liu}}, \bibinfo {author} {\bibfnamefont {Y.}~\bibnamefont {Guan}}, \bibinfo {author} {\bibfnamefont {J.}~\bibnamefont {Shi}}, \bibinfo {author} {\bibfnamefont {F.}~\bibnamefont {Yang}}, \bibinfo {author} {\bibfnamefont {Y.}~\bibnamefont {Bai}}, \bibinfo {author} {\bibfnamefont {H.}~\bibnamefont {Zhang}}, \bibinfo {author} {\bibfnamefont {S.}~\bibnamefont {Fan}}, \bibinfo {author} {\bibfnamefont {W.}~\bibnamefont {Wu}}, \bibinfo {author} {\bibfnamefont {S.}~\bibnamefont {Zhao}},\ and\ \bibinfo {author} {\bibfnamefont {S.}~\bibnamefont {Zhang}},\ }\href {https://doi.org/10.1088/1681-7575/ad023e} {\bibfield  {journal} {\bibinfo  {journal} {Metrologia}\ }\textbf {\bibinfo {volume} {60}},\ \bibinfo {pages} {065012} (\bibinfo {year} {2023})}\BibitemShut {NoStop}%
\bibitem [{\citenamefont {Gerginov}\ \emph {et~al.}(2025)\citenamefont {Gerginov}, \citenamefont {Hoth}, \citenamefont {Heavner}, \citenamefont {Parker}, \citenamefont {Gibble},\ and\ \citenamefont {Sherman}}]{gerginov2025accuracy}%
  \BibitemOpen
  \bibfield  {author} {\bibinfo {author} {\bibfnamefont {V.}~\bibnamefont {Gerginov}}, \bibinfo {author} {\bibfnamefont {G.}~\bibnamefont {Hoth}}, \bibinfo {author} {\bibfnamefont {T.}~\bibnamefont {Heavner}}, \bibinfo {author} {\bibfnamefont {T.}~\bibnamefont {Parker}}, \bibinfo {author} {\bibfnamefont {K.}~\bibnamefont {Gibble}},\ and\ \bibinfo {author} {\bibfnamefont {J.}~\bibnamefont {Sherman}},\ }\href {https://doi.org/10.1088/1681-7575/adc7bd} {\bibfield  {journal} {\bibinfo  {journal} {Metrologia}\ }\textbf {\bibinfo {volume} {62}},\ \bibinfo {pages} {035002} (\bibinfo {year} {2025})}\BibitemShut {NoStop}%
\bibitem [{\citenamefont {Sch\"affner}\ \emph {et~al.}(2024)\citenamefont {Sch\"affner}, \citenamefont {Schreiber}, \citenamefont {Lenz}, \citenamefont {Schlosser},\ and\ \citenamefont {Birkl}}]{Dominik2024Quantum}%
  \BibitemOpen
  \bibfield  {author} {\bibinfo {author} {\bibfnamefont {D.}~\bibnamefont {Sch\"affner}}, \bibinfo {author} {\bibfnamefont {T.}~\bibnamefont {Schreiber}}, \bibinfo {author} {\bibfnamefont {F.}~\bibnamefont {Lenz}}, \bibinfo {author} {\bibfnamefont {M.}~\bibnamefont {Schlosser}},\ and\ \bibinfo {author} {\bibfnamefont {G.}~\bibnamefont {Birkl}},\ }\href {https://doi.org/10.1103/PRXQuantum.5.010311} {\bibfield  {journal} {\bibinfo  {journal} {PRX Quantum}\ }\textbf {\bibinfo {volume} {5}},\ \bibinfo {pages} {010311} (\bibinfo {year} {2024})}\BibitemShut {NoStop}%
\bibitem [{\citenamefont {Ye}\ and\ \citenamefont {Zoller}(2024)}]{Ye2024Quantum}%
  \BibitemOpen
  \bibfield  {author} {\bibinfo {author} {\bibfnamefont {J.}~\bibnamefont {Ye}}\ and\ \bibinfo {author} {\bibfnamefont {P.}~\bibnamefont {Zoller}},\ }\href {https://doi.org/10.1103/PhysRevLett.132.190001} {\bibfield  {journal} {\bibinfo  {journal} {Phys. Rev. Lett.}\ }\textbf {\bibinfo {volume} {132}},\ \bibinfo {pages} {190001} (\bibinfo {year} {2024})}\BibitemShut {NoStop}%
\bibitem [{\citenamefont {Saffman}\ \emph {et~al.}(2010)\citenamefont {Saffman}, \citenamefont {Walker},\ and\ \citenamefont {M\o{}lmer}}]{Saffman2010Quantum}%
  \BibitemOpen
  \bibfield  {author} {\bibinfo {author} {\bibfnamefont {M.}~\bibnamefont {Saffman}}, \bibinfo {author} {\bibfnamefont {T.~G.}\ \bibnamefont {Walker}},\ and\ \bibinfo {author} {\bibfnamefont {K.}~\bibnamefont {M\o{}lmer}},\ }\href {https://doi.org/10.1103/RevModPhys.82.2313} {\bibfield  {journal} {\bibinfo  {journal} {Rev. Mod. Phys.}\ }\textbf {\bibinfo {volume} {82}},\ \bibinfo {pages} {2313} (\bibinfo {year} {2010})}\BibitemShut {NoStop}%
\bibitem [{\citenamefont {Bernien}\ \emph {et~al.}(2017)\citenamefont {Bernien}, \citenamefont {Schwartz}, \citenamefont {Keesling}, \citenamefont {Levine}, \citenamefont {Omran}, \citenamefont {Pichler}, \citenamefont {Choi}, \citenamefont {Zibrov}, \citenamefont {Endres}, \citenamefont {Greiner}, \citenamefont {Vuleti{\'c}},\ and\ \citenamefont {Lukin}}]{Bernien2017Probing}%
  \BibitemOpen
  \bibfield  {author} {\bibinfo {author} {\bibfnamefont {H.}~\bibnamefont {Bernien}}, \bibinfo {author} {\bibfnamefont {S.}~\bibnamefont {Schwartz}}, \bibinfo {author} {\bibfnamefont {A.}~\bibnamefont {Keesling}}, \bibinfo {author} {\bibfnamefont {H.}~\bibnamefont {Levine}}, \bibinfo {author} {\bibfnamefont {A.}~\bibnamefont {Omran}}, \bibinfo {author} {\bibfnamefont {H.}~\bibnamefont {Pichler}}, \bibinfo {author} {\bibfnamefont {S.}~\bibnamefont {Choi}}, \bibinfo {author} {\bibfnamefont {A.~S.}\ \bibnamefont {Zibrov}}, \bibinfo {author} {\bibfnamefont {M.}~\bibnamefont {Endres}}, \bibinfo {author} {\bibfnamefont {M.}~\bibnamefont {Greiner}}, \bibinfo {author} {\bibfnamefont {V.}~\bibnamefont {Vuleti{\'c}}},\ and\ \bibinfo {author} {\bibfnamefont {M.~D.}\ \bibnamefont {Lukin}},\ }\href {https://doi.org/10.1038/nature24622} {\bibfield  {journal} {\bibinfo  {journal} {Nature}\ }\textbf {\bibinfo {volume} {551}},\ \bibinfo {pages} {579} (\bibinfo {year} {2017})}\BibitemShut {NoStop}%
\bibitem [{\citenamefont {Manovitz}\ \emph {et~al.}(2025)\citenamefont {Manovitz}, \citenamefont {Li}, \citenamefont {Ebadi}, \citenamefont {Samajdar}, \citenamefont {Geim}, \citenamefont {Evered}, \citenamefont {Bluvstein}, \citenamefont {Zhou}, \citenamefont {Koyluoglu}, \citenamefont {Feldmeier}, \citenamefont {Dolgirev}, \citenamefont {Maskara}, \citenamefont {Kalinowski}, \citenamefont {Sachdev}, \citenamefont {Huse}, \citenamefont {Greiner}, \citenamefont {Vuletic},\ and\ \citenamefont {Lukin}}]{manovitz2025quantum}%
  \BibitemOpen
  \bibfield  {author} {\bibinfo {author} {\bibfnamefont {T.}~\bibnamefont {Manovitz}}, \bibinfo {author} {\bibfnamefont {S.~H.}\ \bibnamefont {Li}}, \bibinfo {author} {\bibfnamefont {S.}~\bibnamefont {Ebadi}}, \bibinfo {author} {\bibfnamefont {R.}~\bibnamefont {Samajdar}}, \bibinfo {author} {\bibfnamefont {A.~A.}\ \bibnamefont {Geim}}, \bibinfo {author} {\bibfnamefont {S.~J.}\ \bibnamefont {Evered}}, \bibinfo {author} {\bibfnamefont {D.}~\bibnamefont {Bluvstein}}, \bibinfo {author} {\bibfnamefont {H.}~\bibnamefont {Zhou}}, \bibinfo {author} {\bibfnamefont {N.~U.}\ \bibnamefont {Koyluoglu}}, \bibinfo {author} {\bibfnamefont {J.}~\bibnamefont {Feldmeier}}, \bibinfo {author} {\bibfnamefont {P.~E.}\ \bibnamefont {Dolgirev}}, \bibinfo {author} {\bibfnamefont {N.}~\bibnamefont {Maskara}}, \bibinfo {author} {\bibfnamefont {M.}~\bibnamefont {Kalinowski}}, \bibinfo {author} {\bibfnamefont {S.}~\bibnamefont {Sachdev}}, \bibinfo {author} {\bibfnamefont {D.~A.}\ \bibnamefont {Huse}}, \bibinfo {author} {\bibfnamefont
  {M.}~\bibnamefont {Greiner}}, \bibinfo {author} {\bibfnamefont {V.}~\bibnamefont {Vuletic}},\ and\ \bibinfo {author} {\bibfnamefont {M.~D.}\ \bibnamefont {Lukin}},\ }\href {https://doi.org/10.1038/s41586-024-08353-5} {\bibfield  {journal} {\bibinfo  {journal} {Nature}\ }\textbf {\bibinfo {volume} {638}},\ \bibinfo {pages} {86} (\bibinfo {year} {2025})}\BibitemShut {NoStop}%
\bibitem [{\citenamefont {Javanainen}\ and\ \citenamefont {Stenholm}(1981)}]{javanainen1981laser}%
  \BibitemOpen
  \bibfield  {author} {\bibinfo {author} {\bibfnamefont {J.}~\bibnamefont {Javanainen}}\ and\ \bibinfo {author} {\bibfnamefont {S.}~\bibnamefont {Stenholm}},\ }\href {https://doi.org/10.1007/BF00902273} {\bibfield  {journal} {\bibinfo  {journal} {Appl. Phys.}\ }\textbf {\bibinfo {volume} {24}},\ \bibinfo {pages} {151} (\bibinfo {year} {1981})}\BibitemShut {NoStop}%
\bibitem [{\citenamefont {Leibfried}\ \emph {et~al.}(2003)\citenamefont {Leibfried}, \citenamefont {Blatt}, \citenamefont {Monroe},\ and\ \citenamefont {Wineland}}]{leibfried2003quantum}%
  \BibitemOpen
  \bibfield  {author} {\bibinfo {author} {\bibfnamefont {D.}~\bibnamefont {Leibfried}}, \bibinfo {author} {\bibfnamefont {R.}~\bibnamefont {Blatt}}, \bibinfo {author} {\bibfnamefont {C.}~\bibnamefont {Monroe}},\ and\ \bibinfo {author} {\bibfnamefont {D.}~\bibnamefont {Wineland}},\ }\href {https://doi.org/10.1103/RevModPhys.75.281} {\bibfield  {journal} {\bibinfo  {journal} {Rev. Mod. Phys.}\ }\textbf {\bibinfo {volume} {75}},\ \bibinfo {pages} {281} (\bibinfo {year} {2003})}\BibitemShut {NoStop}%
\bibitem [{\citenamefont {Phatak}\ \emph {et~al.}(2024)\citenamefont {Phatak}, \citenamefont {Blodgett}, \citenamefont {Peana}, \citenamefont {Chen},\ and\ \citenamefont {Hood}}]{Phatak2024Generalized}%
  \BibitemOpen
  \bibfield  {author} {\bibinfo {author} {\bibfnamefont {S.~S.}\ \bibnamefont {Phatak}}, \bibinfo {author} {\bibfnamefont {K.~N.}\ \bibnamefont {Blodgett}}, \bibinfo {author} {\bibfnamefont {D.}~\bibnamefont {Peana}}, \bibinfo {author} {\bibfnamefont {M.~R.}\ \bibnamefont {Chen}},\ and\ \bibinfo {author} {\bibfnamefont {J.~D.}\ \bibnamefont {Hood}},\ }\href {https://doi.org/10.1103/PhysRevA.110.043116} {\bibfield  {journal} {\bibinfo  {journal} {Phys. Rev. A}\ }\textbf {\bibinfo {volume} {110}},\ \bibinfo {pages} {043116} (\bibinfo {year} {2024})}\BibitemShut {NoStop}%
\bibitem [{\citenamefont {Diedrich}\ \emph {et~al.}(1989)\citenamefont {Diedrich}, \citenamefont {Bergquist}, \citenamefont {Itano},\ and\ \citenamefont {Wineland}}]{diedrich1989laser}%
  \BibitemOpen
  \bibfield  {author} {\bibinfo {author} {\bibfnamefont {F.}~\bibnamefont {Diedrich}}, \bibinfo {author} {\bibfnamefont {J.~C.}\ \bibnamefont {Bergquist}}, \bibinfo {author} {\bibfnamefont {W.~M.}\ \bibnamefont {Itano}},\ and\ \bibinfo {author} {\bibfnamefont {D.~J.}\ \bibnamefont {Wineland}},\ }\href {https://doi.org/10.1103/PhysRevLett.62.403} {\bibfield  {journal} {\bibinfo  {journal} {Phys. Rev. Lett.}\ }\textbf {\bibinfo {volume} {62}},\ \bibinfo {pages} {403} (\bibinfo {year} {1989})}\BibitemShut {NoStop}%
\bibitem [{\citenamefont {Roos}\ \emph {et~al.}(1999)\citenamefont {Roos}, \citenamefont {Zeiger}, \citenamefont {Rohde}, \citenamefont {N\"agerl}, \citenamefont {Eschner}, \citenamefont {Leibfried}, \citenamefont {Schmidt-Kaler},\ and\ \citenamefont {Blatt}}]{roos1999quantum}%
  \BibitemOpen
  \bibfield  {author} {\bibinfo {author} {\bibfnamefont {C.}~\bibnamefont {Roos}}, \bibinfo {author} {\bibfnamefont {T.}~\bibnamefont {Zeiger}}, \bibinfo {author} {\bibfnamefont {H.}~\bibnamefont {Rohde}}, \bibinfo {author} {\bibfnamefont {H.~C.}\ \bibnamefont {N\"agerl}}, \bibinfo {author} {\bibfnamefont {J.}~\bibnamefont {Eschner}}, \bibinfo {author} {\bibfnamefont {D.}~\bibnamefont {Leibfried}}, \bibinfo {author} {\bibfnamefont {F.}~\bibnamefont {Schmidt-Kaler}},\ and\ \bibinfo {author} {\bibfnamefont {R.}~\bibnamefont {Blatt}},\ }\href {https://doi.org/10.1103/PhysRevLett.83.4713} {\bibfield  {journal} {\bibinfo  {journal} {Phys. Rev. Lett.}\ }\textbf {\bibinfo {volume} {83}},\ \bibinfo {pages} {4713} (\bibinfo {year} {1999})}\BibitemShut {NoStop}%
\bibitem [{\citenamefont {Eschner}\ \emph {et~al.}(2003)\citenamefont {Eschner}, \citenamefont {Morigi}, \citenamefont {Schmidt-Kaler},\ and\ \citenamefont {Blatt}}]{eschner2003laser}%
  \BibitemOpen
  \bibfield  {author} {\bibinfo {author} {\bibfnamefont {J.}~\bibnamefont {Eschner}}, \bibinfo {author} {\bibfnamefont {G.}~\bibnamefont {Morigi}}, \bibinfo {author} {\bibfnamefont {F.}~\bibnamefont {Schmidt-Kaler}},\ and\ \bibinfo {author} {\bibfnamefont {R.}~\bibnamefont {Blatt}},\ }\href {https://doi.org/10.1364/JOSAB.20.001003} {\bibfield  {journal} {\bibinfo  {journal} {J. Opt. Soc. Am. B}\ }\textbf {\bibinfo {volume} {20}},\ \bibinfo {pages} {1003} (\bibinfo {year} {2003})}\BibitemShut {NoStop}%
\bibitem [{\citenamefont {Saskin}\ \emph {et~al.}(2019)\citenamefont {Saskin}, \citenamefont {Wilson}, \citenamefont {Grinkemeyer},\ and\ \citenamefont {Thompson}}]{saskin2019narrow}%
  \BibitemOpen
  \bibfield  {author} {\bibinfo {author} {\bibfnamefont {S.}~\bibnamefont {Saskin}}, \bibinfo {author} {\bibfnamefont {J.~T.}\ \bibnamefont {Wilson}}, \bibinfo {author} {\bibfnamefont {B.}~\bibnamefont {Grinkemeyer}},\ and\ \bibinfo {author} {\bibfnamefont {J.~D.}\ \bibnamefont {Thompson}},\ }\href {https://doi.org/10.1103/PhysRevLett.122.143002} {\bibfield  {journal} {\bibinfo  {journal} {Phys. Rev. Lett.}\ }\textbf {\bibinfo {volume} {122}},\ \bibinfo {pages} {143002} (\bibinfo {year} {2019})}\BibitemShut {NoStop}%
\bibitem [{\citenamefont {Norcia}\ \emph {et~al.}(2018)\citenamefont {Norcia}, \citenamefont {Young},\ and\ \citenamefont {Kaufman}}]{norcia2018microscopic}%
  \BibitemOpen
  \bibfield  {author} {\bibinfo {author} {\bibfnamefont {M.~A.}\ \bibnamefont {Norcia}}, \bibinfo {author} {\bibfnamefont {A.~W.}\ \bibnamefont {Young}},\ and\ \bibinfo {author} {\bibfnamefont {A.~M.}\ \bibnamefont {Kaufman}},\ }\href {https://doi.org/10.1103/PhysRevX.8.041054} {\bibfield  {journal} {\bibinfo  {journal} {Phys. Rev. X}\ }\textbf {\bibinfo {volume} {8}},\ \bibinfo {pages} {041054} (\bibinfo {year} {2018})}\BibitemShut {NoStop}%
\bibitem [{\citenamefont {Norcia}\ \emph {et~al.}(2019)\citenamefont {Norcia}, \citenamefont {Young}, \citenamefont {Eckner}, \citenamefont {Oelker}, \citenamefont {Ye},\ and\ \citenamefont {Kaufman}}]{norcia2019seconds}%
  \BibitemOpen
  \bibfield  {author} {\bibinfo {author} {\bibfnamefont {M.~A.}\ \bibnamefont {Norcia}}, \bibinfo {author} {\bibfnamefont {A.~W.}\ \bibnamefont {Young}}, \bibinfo {author} {\bibfnamefont {W.~J.}\ \bibnamefont {Eckner}}, \bibinfo {author} {\bibfnamefont {E.}~\bibnamefont {Oelker}}, \bibinfo {author} {\bibfnamefont {J.}~\bibnamefont {Ye}},\ and\ \bibinfo {author} {\bibfnamefont {A.~M.}\ \bibnamefont {Kaufman}},\ }\href {https://doi.org/10.1126/science.aay0644} {\bibfield  {journal} {\bibinfo  {journal} {Science}\ }\textbf {\bibinfo {volume} {366}},\ \bibinfo {pages} {93} (\bibinfo {year} {2019})}\BibitemShut {NoStop}%
\bibitem [{\citenamefont {Cooper}\ \emph {et~al.}(2018)\citenamefont {Cooper}, \citenamefont {Covey}, \citenamefont {Madjarov}, \citenamefont {Porsev}, \citenamefont {Safronova},\ and\ \citenamefont {Endres}}]{cooper2018alkaline}%
  \BibitemOpen
  \bibfield  {author} {\bibinfo {author} {\bibfnamefont {A.}~\bibnamefont {Cooper}}, \bibinfo {author} {\bibfnamefont {J.~P.}\ \bibnamefont {Covey}}, \bibinfo {author} {\bibfnamefont {I.~S.}\ \bibnamefont {Madjarov}}, \bibinfo {author} {\bibfnamefont {S.~G.}\ \bibnamefont {Porsev}}, \bibinfo {author} {\bibfnamefont {M.~S.}\ \bibnamefont {Safronova}},\ and\ \bibinfo {author} {\bibfnamefont {M.}~\bibnamefont {Endres}},\ }\href {https://doi.org/10.1103/PhysRevX.8.041055} {\bibfield  {journal} {\bibinfo  {journal} {Phys. Rev. X}\ }\textbf {\bibinfo {volume} {8}},\ \bibinfo {pages} {041055} (\bibinfo {year} {2018})}\BibitemShut {NoStop}%
\bibitem [{\citenamefont {Covey}\ \emph {et~al.}(2019)\citenamefont {Covey}, \citenamefont {Madjarov}, \citenamefont {Cooper},\ and\ \citenamefont {Endres}}]{covey20192000}%
  \BibitemOpen
  \bibfield  {author} {\bibinfo {author} {\bibfnamefont {J.~P.}\ \bibnamefont {Covey}}, \bibinfo {author} {\bibfnamefont {I.~S.}\ \bibnamefont {Madjarov}}, \bibinfo {author} {\bibfnamefont {A.}~\bibnamefont {Cooper}},\ and\ \bibinfo {author} {\bibfnamefont {M.}~\bibnamefont {Endres}},\ }\href {https://doi.org/10.1103/PhysRevLett.122.173201} {\bibfield  {journal} {\bibinfo  {journal} {Phys. Rev. Lett.}\ }\textbf {\bibinfo {volume} {122}},\ \bibinfo {pages} {173201} (\bibinfo {year} {2019})}\BibitemShut {NoStop}%
\bibitem [{\citenamefont {Urech}\ \emph {et~al.}(2022)\citenamefont {Urech}, \citenamefont {Knottnerus}, \citenamefont {Spreeuw},\ and\ \citenamefont {Schreck}}]{urech2022narrow}%
  \BibitemOpen
  \bibfield  {author} {\bibinfo {author} {\bibfnamefont {A.}~\bibnamefont {Urech}}, \bibinfo {author} {\bibfnamefont {I.~H.~A.}\ \bibnamefont {Knottnerus}}, \bibinfo {author} {\bibfnamefont {R.~J.~C.}\ \bibnamefont {Spreeuw}},\ and\ \bibinfo {author} {\bibfnamefont {F.}~\bibnamefont {Schreck}},\ }\href {https://doi.org/10.1103/PhysRevResearch.4.023245} {\bibfield  {journal} {\bibinfo  {journal} {Phys. Rev. Res.}\ }\textbf {\bibinfo {volume} {4}},\ \bibinfo {pages} {023245} (\bibinfo {year} {2022})}\BibitemShut {NoStop}%
\bibitem [{\citenamefont {Ivanov}\ and\ \citenamefont {Gupta}(2011)}]{ivanov2011laser}%
  \BibitemOpen
  \bibfield  {author} {\bibinfo {author} {\bibfnamefont {V.~V.}\ \bibnamefont {Ivanov}}\ and\ \bibinfo {author} {\bibfnamefont {S.}~\bibnamefont {Gupta}},\ }\href {https://doi.org/10.1103/PhysRevA.84.063417} {\bibfield  {journal} {\bibinfo  {journal} {Phys. Rev. A}\ }\textbf {\bibinfo {volume} {84}},\ \bibinfo {pages} {063417} (\bibinfo {year} {2011})}\BibitemShut {NoStop}%
\bibitem [{\citenamefont {Ta\"{\i}eb}\ \emph {et~al.}(1994)\citenamefont {Ta\"{\i}eb}, \citenamefont {Dum}, \citenamefont {Cirac}, \citenamefont {Marte},\ and\ \citenamefont {Zoller}}]{taieb1994coolinga}%
  \BibitemOpen
  \bibfield  {author} {\bibinfo {author} {\bibfnamefont {R.}~\bibnamefont {Ta\"{\i}eb}}, \bibinfo {author} {\bibfnamefont {R.}~\bibnamefont {Dum}}, \bibinfo {author} {\bibfnamefont {J.~I.}\ \bibnamefont {Cirac}}, \bibinfo {author} {\bibfnamefont {P.}~\bibnamefont {Marte}},\ and\ \bibinfo {author} {\bibfnamefont {P.}~\bibnamefont {Zoller}},\ }\href {https://doi.org/10.1103/PhysRevA.49.4876} {\bibfield  {journal} {\bibinfo  {journal} {Phys. Rev. A}\ }\textbf {\bibinfo {volume} {49}},\ \bibinfo {pages} {4876} (\bibinfo {year} {1994})}\BibitemShut {NoStop}%
\bibitem [{\citenamefont {Wineland}\ \emph {et~al.}(1992)\citenamefont {Wineland}, \citenamefont {Dalibard},\ and\ \citenamefont {Cohen-Tannoudji}}]{wineland1992sisyphus}%
  \BibitemOpen
  \bibfield  {author} {\bibinfo {author} {\bibfnamefont {D.~J.}\ \bibnamefont {Wineland}}, \bibinfo {author} {\bibfnamefont {J.}~\bibnamefont {Dalibard}},\ and\ \bibinfo {author} {\bibfnamefont {C.}~\bibnamefont {Cohen-Tannoudji}},\ }\href {https://doi.org/10.1364/JOSAB.9.000032} {\bibfield  {journal} {\bibinfo  {journal} {JOSA B}\ }\textbf {\bibinfo {volume} {9}},\ \bibinfo {pages} {32} (\bibinfo {year} {1992})}\BibitemShut {NoStop}%
\bibitem [{\citenamefont {Graham}\ \emph {et~al.}(2023)\citenamefont {Graham}, \citenamefont {Phuttitarn}, \citenamefont {Chinnarasu}, \citenamefont {Song}, \citenamefont {Poole}, \citenamefont {Jooya}, \citenamefont {Scott}, \citenamefont {Scott}, \citenamefont {Eichler},\ and\ \citenamefont {Saffman}}]{graham2023midcircuit}%
  \BibitemOpen
  \bibfield  {author} {\bibinfo {author} {\bibfnamefont {T.~M.}\ \bibnamefont {Graham}}, \bibinfo {author} {\bibfnamefont {L.}~\bibnamefont {Phuttitarn}}, \bibinfo {author} {\bibfnamefont {R.}~\bibnamefont {Chinnarasu}}, \bibinfo {author} {\bibfnamefont {Y.}~\bibnamefont {Song}}, \bibinfo {author} {\bibfnamefont {C.}~\bibnamefont {Poole}}, \bibinfo {author} {\bibfnamefont {K.}~\bibnamefont {Jooya}}, \bibinfo {author} {\bibfnamefont {J.}~\bibnamefont {Scott}}, \bibinfo {author} {\bibfnamefont {A.}~\bibnamefont {Scott}}, \bibinfo {author} {\bibfnamefont {P.}~\bibnamefont {Eichler}},\ and\ \bibinfo {author} {\bibfnamefont {M.}~\bibnamefont {Saffman}},\ }\href {https://doi.org/10.1103/PhysRevX.13.041051} {\bibfield  {journal} {\bibinfo  {journal} {Phys. Rev. X}\ }\textbf {\bibinfo {volume} {13}},\ \bibinfo {pages} {041051} (\bibinfo {year} {2023})}\BibitemShut {NoStop}%
\bibitem [{\citenamefont {Berto}\ \emph {et~al.}(2021)\citenamefont {Berto}, \citenamefont {Perego}, \citenamefont {Duca},\ and\ \citenamefont {Sias}}]{berto2021prospects}%
  \BibitemOpen
  \bibfield  {author} {\bibinfo {author} {\bibfnamefont {F.}~\bibnamefont {Berto}}, \bibinfo {author} {\bibfnamefont {E.}~\bibnamefont {Perego}}, \bibinfo {author} {\bibfnamefont {L.}~\bibnamefont {Duca}},\ and\ \bibinfo {author} {\bibfnamefont {C.}~\bibnamefont {Sias}},\ }\href {https://doi.org/10.1103/PhysRevResearch.3.043106} {\bibfield  {journal} {\bibinfo  {journal} {Phys. Rev. Res.}\ }\textbf {\bibinfo {volume} {3}},\ \bibinfo {pages} {043106} (\bibinfo {year} {2021})}\BibitemShut {NoStop}%
\bibitem [{\citenamefont {H\"olzl}\ \emph {et~al.}(2023)\citenamefont {H\"olzl}, \citenamefont {G\"otzelmann}, \citenamefont {Wirth}, \citenamefont {Safronova}, \citenamefont {Weber},\ and\ \citenamefont {Meinert}}]{Holzl2023Motional}%
  \BibitemOpen
  \bibfield  {author} {\bibinfo {author} {\bibfnamefont {C.}~\bibnamefont {H\"olzl}}, \bibinfo {author} {\bibfnamefont {A.}~\bibnamefont {G\"otzelmann}}, \bibinfo {author} {\bibfnamefont {M.}~\bibnamefont {Wirth}}, \bibinfo {author} {\bibfnamefont {M.~S.}\ \bibnamefont {Safronova}}, \bibinfo {author} {\bibfnamefont {S.}~\bibnamefont {Weber}},\ and\ \bibinfo {author} {\bibfnamefont {F.}~\bibnamefont {Meinert}},\ }\href {https://doi.org/10.1103/PhysRevResearch.5.033093} {\bibfield  {journal} {\bibinfo  {journal} {Phys. Rev. Res.}\ }\textbf {\bibinfo {volume} {5}},\ \bibinfo {pages} {033093} (\bibinfo {year} {2023})}\BibitemShut {NoStop}%
\bibitem [{\citenamefont {Biagioni}\ \emph {et~al.}(2025)\citenamefont {Biagioni}, \citenamefont {Hofer}, \citenamefont {Bonvalet}, \citenamefont {Bloch}, \citenamefont {Browaeys},\ and\ \citenamefont {Ferrier-Barbut}}]{biagioni2025narrowline}%
  \BibitemOpen
  \bibfield  {author} {\bibinfo {author} {\bibfnamefont {G.}~\bibnamefont {Biagioni}}, \bibinfo {author} {\bibfnamefont {B.}~\bibnamefont {Hofer}}, \bibinfo {author} {\bibfnamefont {N.}~\bibnamefont {Bonvalet}}, \bibinfo {author} {\bibfnamefont {D.}~\bibnamefont {Bloch}}, \bibinfo {author} {\bibfnamefont {A.}~\bibnamefont {Browaeys}},\ and\ \bibinfo {author} {\bibfnamefont {I.}~\bibnamefont {Ferrier-Barbut}},\ }\href@noop {} {\bibfield  {journal} {\bibinfo  {journal} {arXiv preprint arXiv:2505.03456}\ } (\bibinfo {year} {2025})}\BibitemShut {NoStop}%
\bibitem [{\citenamefont {Pucher}\ \emph {et~al.}(2020)\citenamefont {Pucher}, \citenamefont {Schneeweiss}, \citenamefont {Rauschenbeutel},\ and\ \citenamefont {Dareau}}]{pucher2020lifetime}%
  \BibitemOpen
  \bibfield  {author} {\bibinfo {author} {\bibfnamefont {S.}~\bibnamefont {Pucher}}, \bibinfo {author} {\bibfnamefont {P.}~\bibnamefont {Schneeweiss}}, \bibinfo {author} {\bibfnamefont {A.}~\bibnamefont {Rauschenbeutel}},\ and\ \bibinfo {author} {\bibfnamefont {A.}~\bibnamefont {Dareau}},\ }\href {https://doi.org/10.1103/PhysRevA.101.042510} {\bibfield  {journal} {\bibinfo  {journal} {Phys. Rev. A}\ }\textbf {\bibinfo {volume} {101}},\ \bibinfo {pages} {042510} (\bibinfo {year} {2020})}\BibitemShut {NoStop}%
\bibitem [{\citenamefont {Carr}(2014)}]{carr2014improving}%
  \BibitemOpen
  \bibfield  {author} {\bibinfo {author} {\bibfnamefont {A.~W.}\ \bibnamefont {Carr}},\ }\emph {\bibinfo {title} {Improving quantum computation with neutral cesium: Readout and cooling on a quadrupole line, conditions for double magic traps and a novel dissipative entanglement scheme}},\ \href {https://asset.library.wisc.edu/1711.dl/TCN2JSZPJNUDG8M/R/file-ee558.pdf} {\bibinfo {type} {Ph.d.\ thesis}},\ \bibinfo  {school} {The University of Wisconsin–Madison}, \bibinfo {address} {Madison, Wisconsin} (\bibinfo {year} {2014})\BibitemShut {NoStop}%
\bibitem [{\citenamefont {Sharma}\ \emph {et~al.}(2022)\citenamefont {Sharma}, \citenamefont {Kolkowitz},\ and\ \citenamefont {Saffman}}]{sharma2022analysis}%
  \BibitemOpen
  \bibfield  {author} {\bibinfo {author} {\bibfnamefont {A.}~\bibnamefont {Sharma}}, \bibinfo {author} {\bibfnamefont {S.}~\bibnamefont {Kolkowitz}},\ and\ \bibinfo {author} {\bibfnamefont {M.}~\bibnamefont {Saffman}},\ }\href {https://arxiv.org/abs/2203.08708} {\bibinfo {title} {Analysis of a cesium lattice optical clock}} (\bibinfo {year} {2022}),\ \Eprint {https://arxiv.org/abs/2203.08708} {arXiv:2203.08708 [quant-ph]} \BibitemShut {NoStop}%
\bibitem [{\citenamefont {Duspayev}\ \emph {et~al.}(2024)\citenamefont {Duspayev}, \citenamefont {Owens}, \citenamefont {Dash},\ and\ \citenamefont {Raithel}}]{duspayev2024optical}%
  \BibitemOpen
  \bibfield  {author} {\bibinfo {author} {\bibfnamefont {A.}~\bibnamefont {Duspayev}}, \bibinfo {author} {\bibfnamefont {C.}~\bibnamefont {Owens}}, \bibinfo {author} {\bibfnamefont {B.}~\bibnamefont {Dash}},\ and\ \bibinfo {author} {\bibfnamefont {G.}~\bibnamefont {Raithel}},\ }\href {https://doi.org/10.1088/2058-9565/ad77ef} {\bibfield  {journal} {\bibinfo  {journal} {Quantum Sci. Technol.}\ }\textbf {\bibinfo {volume} {9}},\ \bibinfo {pages} {045046} (\bibinfo {year} {2024})}\BibitemShut {NoStop}%
\bibitem [{\citenamefont {Allen}\ \emph {et~al.}(1992)\citenamefont {Allen}, \citenamefont {Beijersbergen}, \citenamefont {Spreeuw},\ and\ \citenamefont {Woerdman}}]{allen1992orbital}%
  \BibitemOpen
  \bibfield  {author} {\bibinfo {author} {\bibfnamefont {L.}~\bibnamefont {Allen}}, \bibinfo {author} {\bibfnamefont {M.~W.}\ \bibnamefont {Beijersbergen}}, \bibinfo {author} {\bibfnamefont {R.~J.~C.}\ \bibnamefont {Spreeuw}},\ and\ \bibinfo {author} {\bibfnamefont {J.~P.}\ \bibnamefont {Woerdman}},\ }\href {https://doi.org/10.1103/PhysRevA.45.8185} {\bibfield  {journal} {\bibinfo  {journal} {Phys. Rev. A}\ }\textbf {\bibinfo {volume} {45}},\ \bibinfo {pages} {8185} (\bibinfo {year} {1992})}\BibitemShut {NoStop}%
\bibitem [{\citenamefont {He}\ \emph {et~al.}(1995)\citenamefont {He}, \citenamefont {Friese}, \citenamefont {Heckenberg},\ and\ \citenamefont {Rubinsztein-Dunlop}}]{he1995direct}%
  \BibitemOpen
  \bibfield  {author} {\bibinfo {author} {\bibfnamefont {H.}~\bibnamefont {He}}, \bibinfo {author} {\bibfnamefont {M.~E.~J.}\ \bibnamefont {Friese}}, \bibinfo {author} {\bibfnamefont {N.~R.}\ \bibnamefont {Heckenberg}},\ and\ \bibinfo {author} {\bibfnamefont {H.}~\bibnamefont {Rubinsztein-Dunlop}},\ }\href {https://doi.org/10.1103/PhysRevLett.75.826} {\bibfield  {journal} {\bibinfo  {journal} {Phys. Rev. Lett.}\ }\textbf {\bibinfo {volume} {75}},\ \bibinfo {pages} {826} (\bibinfo {year} {1995})}\BibitemShut {NoStop}%
\bibitem [{\citenamefont {Padgett}\ and\ \citenamefont {Bowman}(2011)}]{padgett2011tweezers}%
  \BibitemOpen
  \bibfield  {author} {\bibinfo {author} {\bibfnamefont {M.}~\bibnamefont {Padgett}}\ and\ \bibinfo {author} {\bibfnamefont {R.}~\bibnamefont {Bowman}},\ }\href {https://doi.org/10.1038/nphoton.2011.81} {\bibfield  {journal} {\bibinfo  {journal} {Nat. Photonics}\ }\textbf {\bibinfo {volume} {5}},\ \bibinfo {pages} {343} (\bibinfo {year} {2011})}\BibitemShut {NoStop}%
\bibitem [{\citenamefont {Stilgoe}\ \emph {et~al.}(2022)\citenamefont {Stilgoe}, \citenamefont {Nieminen},\ and\ \citenamefont {Rubinsztein-Dunlop}}]{stilgoe2022controlled}%
  \BibitemOpen
  \bibfield  {author} {\bibinfo {author} {\bibfnamefont {A.~B.}\ \bibnamefont {Stilgoe}}, \bibinfo {author} {\bibfnamefont {T.~A.}\ \bibnamefont {Nieminen}},\ and\ \bibinfo {author} {\bibfnamefont {H.}~\bibnamefont {Rubinsztein-Dunlop}},\ }\href {https://doi.org/10.1038/s41566-022-00983-3} {\bibfield  {journal} {\bibinfo  {journal} {Nat. Photonics}\ }\textbf {\bibinfo {volume} {16}},\ \bibinfo {pages} {346} (\bibinfo {year} {2022})}\BibitemShut {NoStop}%
\bibitem [{\citenamefont {Andersen}\ \emph {et~al.}(2006)\citenamefont {Andersen}, \citenamefont {Ryu}, \citenamefont {Clad\'e}, \citenamefont {Natarajan}, \citenamefont {Vaziri}, \citenamefont {Helmerson},\ and\ \citenamefont {Phillips}}]{andersen2006quantized}%
  \BibitemOpen
  \bibfield  {author} {\bibinfo {author} {\bibfnamefont {M.~F.}\ \bibnamefont {Andersen}}, \bibinfo {author} {\bibfnamefont {C.}~\bibnamefont {Ryu}}, \bibinfo {author} {\bibfnamefont {P.}~\bibnamefont {Clad\'e}}, \bibinfo {author} {\bibfnamefont {V.}~\bibnamefont {Natarajan}}, \bibinfo {author} {\bibfnamefont {A.}~\bibnamefont {Vaziri}}, \bibinfo {author} {\bibfnamefont {K.}~\bibnamefont {Helmerson}},\ and\ \bibinfo {author} {\bibfnamefont {W.~D.}\ \bibnamefont {Phillips}},\ }\href {https://doi.org/10.1103/PhysRevLett.97.170406} {\bibfield  {journal} {\bibinfo  {journal} {Phys. Rev. Lett.}\ }\textbf {\bibinfo {volume} {97}},\ \bibinfo {pages} {170406} (\bibinfo {year} {2006})}\BibitemShut {NoStop}%
\bibitem [{\citenamefont {Verde}\ \emph {et~al.}(2023)\citenamefont {Verde}, \citenamefont {Schmiegelow}, \citenamefont {Poschinger},\ and\ \citenamefont {Schmidt-Kaler}}]{verde2023trapped}%
  \BibitemOpen
  \bibfield  {author} {\bibinfo {author} {\bibfnamefont {M.}~\bibnamefont {Verde}}, \bibinfo {author} {\bibfnamefont {C.~T.}\ \bibnamefont {Schmiegelow}}, \bibinfo {author} {\bibfnamefont {U.}~\bibnamefont {Poschinger}},\ and\ \bibinfo {author} {\bibfnamefont {F.}~\bibnamefont {Schmidt-Kaler}},\ }\href {https://doi.org/10.1038/s41598-023-48589-1} {\bibfield  {journal} {\bibinfo  {journal} {Sci. Rep.}\ }\textbf {\bibinfo {volume} {13}},\ \bibinfo {pages} {21283} (\bibinfo {year} {2023})}\BibitemShut {NoStop}%
\bibitem [{\citenamefont {Afanasev}\ \emph {et~al.}(2018)\citenamefont {Afanasev}, \citenamefont {Carlson}, \citenamefont {Schmiegelow}, \citenamefont {Schulz}, \citenamefont {Schmidt-Kaler},\ and\ \citenamefont {Solyanik}}]{afanasev2018experimental}%
  \BibitemOpen
  \bibfield  {author} {\bibinfo {author} {\bibfnamefont {A.}~\bibnamefont {Afanasev}}, \bibinfo {author} {\bibfnamefont {C.~E.}\ \bibnamefont {Carlson}}, \bibinfo {author} {\bibfnamefont {C.~T.}\ \bibnamefont {Schmiegelow}}, \bibinfo {author} {\bibfnamefont {J.}~\bibnamefont {Schulz}}, \bibinfo {author} {\bibfnamefont {F.}~\bibnamefont {Schmidt-Kaler}},\ and\ \bibinfo {author} {\bibfnamefont {M.}~\bibnamefont {Solyanik}},\ }\href {https://doi.org/10.1088/1367-2630/aaa63d} {\bibfield  {journal} {\bibinfo  {journal} {New J. Phys.}\ }\textbf {\bibinfo {volume} {20}},\ \bibinfo {pages} {023032} (\bibinfo {year} {2018})}\BibitemShut {NoStop}%
\bibitem [{\citenamefont {Stopp}\ \emph {et~al.}(2022)\citenamefont {Stopp}, \citenamefont {Verde}, \citenamefont {Katz}, \citenamefont {Drechsler}, \citenamefont {Schmiegelow},\ and\ \citenamefont {Schmidt-Kaler}}]{stopp2022coherent}%
  \BibitemOpen
  \bibfield  {author} {\bibinfo {author} {\bibfnamefont {F.}~\bibnamefont {Stopp}}, \bibinfo {author} {\bibfnamefont {M.}~\bibnamefont {Verde}}, \bibinfo {author} {\bibfnamefont {M.}~\bibnamefont {Katz}}, \bibinfo {author} {\bibfnamefont {M.}~\bibnamefont {Drechsler}}, \bibinfo {author} {\bibfnamefont {C.~T.}\ \bibnamefont {Schmiegelow}},\ and\ \bibinfo {author} {\bibfnamefont {F.}~\bibnamefont {Schmidt-Kaler}},\ }\href {https://doi.org/10.1103/PhysRevLett.129.263603} {\bibfield  {journal} {\bibinfo  {journal} {Phys. Rev. Lett.}\ }\textbf {\bibinfo {volume} {129}},\ \bibinfo {pages} {263603} (\bibinfo {year} {2022})}\BibitemShut {NoStop}%
\bibitem [{\citenamefont {Quinteiro}\ \emph {et~al.}(2017)\citenamefont {Quinteiro}, \citenamefont {Schmidt-Kaler},\ and\ \citenamefont {Schmiegelow}}]{quinteiro2017twisted}%
  \BibitemOpen
  \bibfield  {author} {\bibinfo {author} {\bibfnamefont {G.~F.}\ \bibnamefont {Quinteiro}}, \bibinfo {author} {\bibfnamefont {F.}~\bibnamefont {Schmidt-Kaler}},\ and\ \bibinfo {author} {\bibfnamefont {C.~T.}\ \bibnamefont {Schmiegelow}},\ }\href {https://doi.org/10.1103/PhysRevLett.119.253203} {\bibfield  {journal} {\bibinfo  {journal} {Phys. Rev. Lett.}\ }\textbf {\bibinfo {volume} {119}},\ \bibinfo {pages} {253203} (\bibinfo {year} {2017})}\BibitemShut {NoStop}%
\bibitem [{\citenamefont {Quinteiro}\ \emph {et~al.}(2020)\citenamefont {Quinteiro}, \citenamefont {Schmiegelow},\ and\ \citenamefont {Schmidt-Kaler}}]{quinteiro2020paraxial}%
  \BibitemOpen
  \bibfield  {author} {\bibinfo {author} {\bibfnamefont {G.~F.}\ \bibnamefont {Quinteiro}}, \bibinfo {author} {\bibfnamefont {C.~T.}\ \bibnamefont {Schmiegelow}},\ and\ \bibinfo {author} {\bibfnamefont {F.}~\bibnamefont {Schmidt-Kaler}},\ }\href {https://arxiv.org/abs/2004.00040} {\bibfield  {journal} {\bibinfo  {journal} {arXiv preprint arXiv:2004.00040}\ } (\bibinfo {year} {2020})},\ \Eprint {https://arxiv.org/abs/2004.00040} {arXiv:2004.00040 [physics.optics]} \BibitemShut {NoStop}%
\bibitem [{\citenamefont {Lange}\ \emph {et~al.}(2022)\citenamefont {Lange}, \citenamefont {Huntemann}, \citenamefont {Peshkov}, \citenamefont {Surzhykov},\ and\ \citenamefont {Peik}}]{lange2022excitation}%
  \BibitemOpen
  \bibfield  {author} {\bibinfo {author} {\bibfnamefont {R.}~\bibnamefont {Lange}}, \bibinfo {author} {\bibfnamefont {N.}~\bibnamefont {Huntemann}}, \bibinfo {author} {\bibfnamefont {A.~A.}\ \bibnamefont {Peshkov}}, \bibinfo {author} {\bibfnamefont {A.}~\bibnamefont {Surzhykov}},\ and\ \bibinfo {author} {\bibfnamefont {E.}~\bibnamefont {Peik}},\ }\href {https://doi.org/10.1103/PhysRevLett.129.253901} {\bibfield  {journal} {\bibinfo  {journal} {Phys. Rev. Lett.}\ }\textbf {\bibinfo {volume} {129}},\ \bibinfo {pages} {253901} (\bibinfo {year} {2022})}\BibitemShut {NoStop}%
\bibitem [{\citenamefont {Levine}\ \emph {et~al.}(2018)\citenamefont {Levine}, \citenamefont {Keesling}, \citenamefont {Omran}, \citenamefont {Bernien}, \citenamefont {Schwartz}, \citenamefont {Zibrov}, \citenamefont {Endres}, \citenamefont {Greiner}, \citenamefont {Vuleti\ifmmode~\acute{c}\else \'{c}\fi{}},\ and\ \citenamefont {Lukin}}]{levine2018high}%
  \BibitemOpen
  \bibfield  {author} {\bibinfo {author} {\bibfnamefont {H.}~\bibnamefont {Levine}}, \bibinfo {author} {\bibfnamefont {A.}~\bibnamefont {Keesling}}, \bibinfo {author} {\bibfnamefont {A.}~\bibnamefont {Omran}}, \bibinfo {author} {\bibfnamefont {H.}~\bibnamefont {Bernien}}, \bibinfo {author} {\bibfnamefont {S.}~\bibnamefont {Schwartz}}, \bibinfo {author} {\bibfnamefont {A.~S.}\ \bibnamefont {Zibrov}}, \bibinfo {author} {\bibfnamefont {M.}~\bibnamefont {Endres}}, \bibinfo {author} {\bibfnamefont {M.}~\bibnamefont {Greiner}}, \bibinfo {author} {\bibfnamefont {V.}~\bibnamefont {Vuleti\ifmmode~\acute{c}\else \'{c}\fi{}}},\ and\ \bibinfo {author} {\bibfnamefont {M.~D.}\ \bibnamefont {Lukin}},\ }\href {https://doi.org/10.1103/PhysRevLett.121.123603} {\bibfield  {journal} {\bibinfo  {journal} {Phys. Rev. Lett.}\ }\textbf {\bibinfo {volume} {121}},\ \bibinfo {pages} {123603} (\bibinfo {year} {2018})}\BibitemShut {NoStop}%
\bibitem [{\citenamefont {Liu}\ \emph {et~al.}(2019)\citenamefont {Liu}, \citenamefont {Hood}, \citenamefont {Yu}, \citenamefont {Zhang}, \citenamefont {Wang}, \citenamefont {Lin}, \citenamefont {Rosenband},\ and\ \citenamefont {Ni}}]{liu2019molecular}%
  \BibitemOpen
  \bibfield  {author} {\bibinfo {author} {\bibfnamefont {L.~R.}\ \bibnamefont {Liu}}, \bibinfo {author} {\bibfnamefont {J.~D.}\ \bibnamefont {Hood}}, \bibinfo {author} {\bibfnamefont {Y.}~\bibnamefont {Yu}}, \bibinfo {author} {\bibfnamefont {J.~T.}\ \bibnamefont {Zhang}}, \bibinfo {author} {\bibfnamefont {K.}~\bibnamefont {Wang}}, \bibinfo {author} {\bibfnamefont {Y.-W.}\ \bibnamefont {Lin}}, \bibinfo {author} {\bibfnamefont {T.}~\bibnamefont {Rosenband}},\ and\ \bibinfo {author} {\bibfnamefont {K.-K.}\ \bibnamefont {Ni}},\ }\href {https://doi.org/10.1103/PhysRevX.9.021039} {\bibfield  {journal} {\bibinfo  {journal} {Phys. Rev. X}\ }\textbf {\bibinfo {volume} {9}},\ \bibinfo {pages} {021039} (\bibinfo {year} {2019})}\BibitemShut {NoStop}%
\bibitem [{\citenamefont {Cairncross}\ \emph {et~al.}(2021)\citenamefont {Cairncross}, \citenamefont {Zhang}, \citenamefont {Picard}, \citenamefont {Yu}, \citenamefont {Wang},\ and\ \citenamefont {Ni}}]{zhang2020forming}%
  \BibitemOpen
  \bibfield  {author} {\bibinfo {author} {\bibfnamefont {W.~B.}\ \bibnamefont {Cairncross}}, \bibinfo {author} {\bibfnamefont {J.~T.}\ \bibnamefont {Zhang}}, \bibinfo {author} {\bibfnamefont {L.~R.~B.}\ \bibnamefont {Picard}}, \bibinfo {author} {\bibfnamefont {Y.}~\bibnamefont {Yu}}, \bibinfo {author} {\bibfnamefont {K.}~\bibnamefont {Wang}},\ and\ \bibinfo {author} {\bibfnamefont {K.-K.}\ \bibnamefont {Ni}},\ }\href {https://doi.org/10.1103/PhysRevLett.126.123402} {\bibfield  {journal} {\bibinfo  {journal} {Phys. Rev. Lett.}\ }\textbf {\bibinfo {volume} {126}},\ \bibinfo {pages} {123402} (\bibinfo {year} {2021})}\BibitemShut {NoStop}%
\bibitem [{\citenamefont {Tiecke}\ \emph {et~al.}(2014)\citenamefont {Tiecke}, \citenamefont {Thompson}, \citenamefont {de~Leon}, \citenamefont {Liu}, \citenamefont {Vuleti{\'c}},\ and\ \citenamefont {Lukin}}]{tiecke2014nanophotonic}%
  \BibitemOpen
  \bibfield  {author} {\bibinfo {author} {\bibfnamefont {T.~G.}\ \bibnamefont {Tiecke}}, \bibinfo {author} {\bibfnamefont {J.~D.}\ \bibnamefont {Thompson}}, \bibinfo {author} {\bibfnamefont {N.~P.}\ \bibnamefont {de~Leon}}, \bibinfo {author} {\bibfnamefont {L.~R.}\ \bibnamefont {Liu}}, \bibinfo {author} {\bibfnamefont {V.}~\bibnamefont {Vuleti{\'c}}},\ and\ \bibinfo {author} {\bibfnamefont {M.~D.}\ \bibnamefont {Lukin}},\ }\href {https://doi.org/10.1038/nature13188} {\bibfield  {journal} {\bibinfo  {journal} {Nature}\ }\textbf {\bibinfo {volume} {508}},\ \bibinfo {pages} {241} (\bibinfo {year} {2014})}\BibitemShut {NoStop}%
\bibitem [{\citenamefont {Hood}\ \emph {et~al.}(2016)\citenamefont {Hood}, \citenamefont {Goban}, \citenamefont {Asenjo-Garcia}, \citenamefont {Lu}, \citenamefont {Yu}, \citenamefont {Chang},\ and\ \citenamefont {Kimble}}]{hood2016atomatom}%
  \BibitemOpen
  \bibfield  {author} {\bibinfo {author} {\bibfnamefont {J.~D.}\ \bibnamefont {Hood}}, \bibinfo {author} {\bibfnamefont {A.}~\bibnamefont {Goban}}, \bibinfo {author} {\bibfnamefont {A.}~\bibnamefont {Asenjo-Garcia}}, \bibinfo {author} {\bibfnamefont {M.}~\bibnamefont {Lu}}, \bibinfo {author} {\bibfnamefont {S.~P.}\ \bibnamefont {Yu}}, \bibinfo {author} {\bibfnamefont {D.~E.}\ \bibnamefont {Chang}},\ and\ \bibinfo {author} {\bibfnamefont {H.~J.}\ \bibnamefont {Kimble}},\ }\href {https://doi.org/10.1073/pnas.1603788113} {\bibfield  {journal} {\bibinfo  {journal} {Proc. Natl. Acad. Sci. U.S.A.}\ }\textbf {\bibinfo {volume} {113}},\ \bibinfo {pages} {10507} (\bibinfo {year} {2016})}\BibitemShut {NoStop}%
\bibitem [{\citenamefont {Zhou}\ \emph {et~al.}(2024)\citenamefont {Zhou}, \citenamefont {Tamura}, \citenamefont {Chang},\ and\ \citenamefont {Hung}}]{zhou2024trapped}%
  \BibitemOpen
  \bibfield  {author} {\bibinfo {author} {\bibfnamefont {X.}~\bibnamefont {Zhou}}, \bibinfo {author} {\bibfnamefont {H.}~\bibnamefont {Tamura}}, \bibinfo {author} {\bibfnamefont {T.-H.}\ \bibnamefont {Chang}},\ and\ \bibinfo {author} {\bibfnamefont {C.-L.}\ \bibnamefont {Hung}},\ }\href {https://doi.org/10.1103/PhysRevX.14.031004} {\bibfield  {journal} {\bibinfo  {journal} {Phys. Rev. X}\ }\textbf {\bibinfo {volume} {14}},\ \bibinfo {pages} {031004} (\bibinfo {year} {2024})}\BibitemShut {NoStop}%
\bibitem [{\citenamefont {Tojo}\ \emph {et~al.}(2004)\citenamefont {Tojo}, \citenamefont {Hasuo},\ and\ \citenamefont {Fujimoto}}]{tojo2004absorption}%
  \BibitemOpen
  \bibfield  {author} {\bibinfo {author} {\bibfnamefont {S.}~\bibnamefont {Tojo}}, \bibinfo {author} {\bibfnamefont {M.}~\bibnamefont {Hasuo}},\ and\ \bibinfo {author} {\bibfnamefont {T.}~\bibnamefont {Fujimoto}},\ }\href {https://doi.org/10.1103/PhysRevLett.92.053001} {\bibfield  {journal} {\bibinfo  {journal} {Phys. Rev. Lett.}\ }\textbf {\bibinfo {volume} {92}},\ \bibinfo {pages} {053001} (\bibinfo {year} {2004})}\BibitemShut {NoStop}%
\bibitem [{\citenamefont {Chan}\ \emph {et~al.}(2016)\citenamefont {Chan}, \citenamefont {Aljunid}, \citenamefont {Zheludev}, \citenamefont {Wilkowski},\ and\ \citenamefont {Ducloy}}]{chan2016doppler}%
  \BibitemOpen
  \bibfield  {author} {\bibinfo {author} {\bibfnamefont {E.~A.}\ \bibnamefont {Chan}}, \bibinfo {author} {\bibfnamefont {S.~A.}\ \bibnamefont {Aljunid}}, \bibinfo {author} {\bibfnamefont {N.~I.}\ \bibnamefont {Zheludev}}, \bibinfo {author} {\bibfnamefont {D.}~\bibnamefont {Wilkowski}},\ and\ \bibinfo {author} {\bibfnamefont {M.}~\bibnamefont {Ducloy}},\ }\href {https://doi.org/10.1364/OL.41.002005} {\bibfield  {journal} {\bibinfo  {journal} {Opt. Lett.}\ }\textbf {\bibinfo {volume} {41}},\ \bibinfo {pages} {2005} (\bibinfo {year} {2016})}\BibitemShut {NoStop}%
\bibitem [{\citenamefont {Murphree}(2020)}]{murphree2020quadrupole}%
  \BibitemOpen
  \bibfield  {author} {\bibinfo {author} {\bibfnamefont {J.}~\bibnamefont {Murphree}},\ }\emph {\bibinfo {title} {Quadrupole Raman Transitions Driven by Optical Vortex Beams in Ultracold Atomic Clouds}},\ \href {https://urresearch.rochester.edu/institutionalPublicationPublicView.action?institutionalItemId=35546} {\bibinfo {type} {Ph.d.\ thesis}},\ \bibinfo  {school} {University of Rochester}, \bibinfo {address} {Rochester, New York} (\bibinfo {year} {2020}),\ \bibinfo {note} {supervised by Professor Nicholas P. Bigelow}\BibitemShut {NoStop}%
\bibitem [{\citenamefont {Bougouffa}\ and\ \citenamefont {Babiker}(2020)}]{bougouffa2020qadrupole}%
  \BibitemOpen
  \bibfield  {author} {\bibinfo {author} {\bibfnamefont {S.}~\bibnamefont {Bougouffa}}\ and\ \bibinfo {author} {\bibfnamefont {M.}~\bibnamefont {Babiker}},\ }\href {https://doi.org/10.1103/PhysRevA.102.063706} {\bibfield  {journal} {\bibinfo  {journal} {Phys. Rev. A}\ }\textbf {\bibinfo {volume} {102}},\ \bibinfo {pages} {063706} (\bibinfo {year} {2020})}\BibitemShut {NoStop}%
\bibitem [{\citenamefont {Gallagher}\ \emph {et~al.}(2025)\citenamefont {Gallagher}, \citenamefont {Mazzanti}, \citenamefont {Ackerman}, \citenamefont {Safavi-Naini}, \citenamefont {Gerritsma},\ and\ \citenamefont {Spreeuw}}]{gallagher2025nonparaxial}%
  \BibitemOpen
  \bibfield  {author} {\bibinfo {author} {\bibfnamefont {L.~P.~H.}\ \bibnamefont {Gallagher}}, \bibinfo {author} {\bibfnamefont {M.}~\bibnamefont {Mazzanti}}, \bibinfo {author} {\bibfnamefont {Z.~E.~D.}\ \bibnamefont {Ackerman}}, \bibinfo {author} {\bibfnamefont {A.}~\bibnamefont {Safavi-Naini}}, \bibinfo {author} {\bibfnamefont {R.}~\bibnamefont {Gerritsma}},\ and\ \bibinfo {author} {\bibfnamefont {R.~J.~C.}\ \bibnamefont {Spreeuw}},\ }\href {https://arxiv.org/abs/2502.19345} {\bibinfo {title} {Non-paraxial effects on laser-qubit interactions}} (\bibinfo {year} {2025}),\ \Eprint {https://arxiv.org/abs/2502.19345} {arXiv:2502.19345 [quant-ph]} \BibitemShut {NoStop}%
\bibitem [{\citenamefont {Carr}\ and\ \citenamefont {Saffman}(2016)}]{carr2016doubly}%
  \BibitemOpen
  \bibfield  {author} {\bibinfo {author} {\bibfnamefont {A.~W.}\ \bibnamefont {Carr}}\ and\ \bibinfo {author} {\bibfnamefont {M.}~\bibnamefont {Saffman}},\ }\href {https://doi.org/10.1103/PhysRevLett.117.150801} {\bibfield  {journal} {\bibinfo  {journal} {Phys. Rev. Lett.}\ }\textbf {\bibinfo {volume} {117}},\ \bibinfo {pages} {150801} (\bibinfo {year} {2016})}\BibitemShut {NoStop}%
\bibitem [{\citenamefont {Barakhshan}\ \emph {et~al.}(2022)\citenamefont {Barakhshan}, \citenamefont {Marrs}, \citenamefont {Bhosale}, \citenamefont {Arora}, \citenamefont {Eigenmann},\ and\ \citenamefont {Safronova}}]{SafronovaDatabase2022}%
  \BibitemOpen
  \bibfield  {author} {\bibinfo {author} {\bibfnamefont {P.}~\bibnamefont {Barakhshan}}, \bibinfo {author} {\bibfnamefont {A.}~\bibnamefont {Marrs}}, \bibinfo {author} {\bibfnamefont {A.}~\bibnamefont {Bhosale}}, \bibinfo {author} {\bibfnamefont {B.}~\bibnamefont {Arora}}, \bibinfo {author} {\bibfnamefont {R.}~\bibnamefont {Eigenmann}},\ and\ \bibinfo {author} {\bibfnamefont {M.~S.}\ \bibnamefont {Safronova}},\ }\href@noop {} {\bibinfo {title} {Portal for high-precision atomic data and computation (version 2.0)}},\ \bibinfo {howpublished} {\url{https://www.udel.edu/atom/}} (\bibinfo {year} {2022})\BibitemShut {NoStop}%
\bibitem [{\citenamefont {Kien}\ \emph {et~al.}(2013)\citenamefont {Kien}, \citenamefont {Schneeweiss},\ and\ \citenamefont {Rauschenbeutel}}]{LeKien2013Dynamical}%
  \BibitemOpen
  \bibfield  {author} {\bibinfo {author} {\bibfnamefont {F.~L.}\ \bibnamefont {Kien}}, \bibinfo {author} {\bibfnamefont {P.}~\bibnamefont {Schneeweiss}},\ and\ \bibinfo {author} {\bibfnamefont {A.}~\bibnamefont {Rauschenbeutel}},\ }\href {https://doi.org/10.1140/epjd/e2013-30729-x} {\bibfield  {journal} {\bibinfo  {journal} {Eur. Phys. J. D}\ }\textbf {\bibinfo {volume} {67}},\ \bibinfo {pages} {92} (\bibinfo {year} {2013})}\BibitemShut {NoStop}%
\bibitem [{\citenamefont {Schaefer}\ \emph {et~al.}(2007)\citenamefont {Schaefer}, \citenamefont {Collett}, \citenamefont {Smyth}, \citenamefont {Barrett},\ and\ \citenamefont {Fraher}}]{schaefer2007measuring}%
  \BibitemOpen
  \bibfield  {author} {\bibinfo {author} {\bibfnamefont {B.}~\bibnamefont {Schaefer}}, \bibinfo {author} {\bibfnamefont {E.}~\bibnamefont {Collett}}, \bibinfo {author} {\bibfnamefont {R.}~\bibnamefont {Smyth}}, \bibinfo {author} {\bibfnamefont {D.}~\bibnamefont {Barrett}},\ and\ \bibinfo {author} {\bibfnamefont {B.}~\bibnamefont {Fraher}},\ }\href {https://doi.org/10.1119/1.2386162} {\bibfield  {journal} {\bibinfo  {journal} {Am. J. Phys.}\ }\textbf {\bibinfo {volume} {75}},\ \bibinfo {pages} {163} (\bibinfo {year} {2007})}\BibitemShut {NoStop}%
\bibitem [{\citenamefont {DiBerardino}\ \emph {et~al.}(1998)\citenamefont {DiBerardino}, \citenamefont {Tanner},\ and\ \citenamefont {Sieradzan}}]{diberardino1998lifetime}%
  \BibitemOpen
  \bibfield  {author} {\bibinfo {author} {\bibfnamefont {D.}~\bibnamefont {DiBerardino}}, \bibinfo {author} {\bibfnamefont {C.~E.}\ \bibnamefont {Tanner}},\ and\ \bibinfo {author} {\bibfnamefont {A.}~\bibnamefont {Sieradzan}},\ }\href {https://doi.org/10.1103/PhysRevA.57.4204} {\bibfield  {journal} {\bibinfo  {journal} {Phys. Rev. A}\ }\textbf {\bibinfo {volume} {57}},\ \bibinfo {pages} {4204} (\bibinfo {year} {1998})}\BibitemShut {NoStop}%
\bibitem [{\citenamefont {Hoeling}\ \emph {et~al.}(1996)\citenamefont {Hoeling}, \citenamefont {Yeh}, \citenamefont {Takekoshi},\ and\ \citenamefont {Knize}}]{Hoeling1996Lifetime}%
  \BibitemOpen
  \bibfield  {author} {\bibinfo {author} {\bibfnamefont {B.}~\bibnamefont {Hoeling}}, \bibinfo {author} {\bibfnamefont {J.~R.}\ \bibnamefont {Yeh}}, \bibinfo {author} {\bibfnamefont {T.}~\bibnamefont {Takekoshi}},\ and\ \bibinfo {author} {\bibfnamefont {R.~J.}\ \bibnamefont {Knize}},\ }\href {https://doi.org/10.1364/OL.21.000074} {\bibfield  {journal} {\bibinfo  {journal} {Opt. Lett.}\ }\textbf {\bibinfo {volume} {21}},\ \bibinfo {pages} {74} (\bibinfo {year} {1996})}\BibitemShut {NoStop}%
\bibitem [{\citenamefont {Hu}\ \emph {et~al.}(2020)\citenamefont {Hu}, \citenamefont {Wang}, \citenamefont {Wang}, \citenamefont {Ji}, \citenamefont {Zhang}, \citenamefont {Li}, \citenamefont {Zhu}, \citenamefont {Wu},\ and\ \citenamefont {Chu}}]{hu2020efficient}%
  \BibitemOpen
  \bibfield  {author} {\bibinfo {author} {\bibfnamefont {Y.}~\bibnamefont {Hu}}, \bibinfo {author} {\bibfnamefont {Z.}~\bibnamefont {Wang}}, \bibinfo {author} {\bibfnamefont {X.}~\bibnamefont {Wang}}, \bibinfo {author} {\bibfnamefont {S.}~\bibnamefont {Ji}}, \bibinfo {author} {\bibfnamefont {C.}~\bibnamefont {Zhang}}, \bibinfo {author} {\bibfnamefont {J.}~\bibnamefont {Li}}, \bibinfo {author} {\bibfnamefont {W.}~\bibnamefont {Zhu}}, \bibinfo {author} {\bibfnamefont {D.}~\bibnamefont {Wu}},\ and\ \bibinfo {author} {\bibfnamefont {J.}~\bibnamefont {Chu}},\ }\href {https://doi.org/10.1038/s41377-020-00362-z} {\bibfield  {journal} {\bibinfo  {journal} {Light: Science \& Applications}\ }\textbf {\bibinfo {volume} {9}},\ \bibinfo {pages} {119} (\bibinfo {year} {2020})}\BibitemShut {NoStop}%
\bibitem [{\citenamefont {Toh}\ \emph {et~al.}(2019)\citenamefont {Toh}, \citenamefont {Chalus}, \citenamefont {Burgess}, \citenamefont {Damitz}, \citenamefont {Imany}, \citenamefont {Leaird}, \citenamefont {Weiner}, \citenamefont {Tanner},\ and\ \citenamefont {Elliott}}]{Toh2019Measurement}%
  \BibitemOpen
  \bibfield  {author} {\bibinfo {author} {\bibfnamefont {G.}~\bibnamefont {Toh}}, \bibinfo {author} {\bibfnamefont {N.}~\bibnamefont {Chalus}}, \bibinfo {author} {\bibfnamefont {A.}~\bibnamefont {Burgess}}, \bibinfo {author} {\bibfnamefont {A.}~\bibnamefont {Damitz}}, \bibinfo {author} {\bibfnamefont {P.}~\bibnamefont {Imany}}, \bibinfo {author} {\bibfnamefont {D.~E.}\ \bibnamefont {Leaird}}, \bibinfo {author} {\bibfnamefont {A.~M.}\ \bibnamefont {Weiner}}, \bibinfo {author} {\bibfnamefont {C.~E.}\ \bibnamefont {Tanner}},\ and\ \bibinfo {author} {\bibfnamefont {D.~S.}\ \bibnamefont {Elliott}},\ }\href {https://doi.org/10.1103/PhysRevA.100.052507} {\bibfield  {journal} {\bibinfo  {journal} {Phys. Rev. A}\ }\textbf {\bibinfo {volume} {100}},\ \bibinfo {pages} {052507} (\bibinfo {year} {2019})}\BibitemShut {NoStop}%
\bibitem [{\citenamefont {Duarte}\ \emph {et~al.}(2011)\citenamefont {Duarte}, \citenamefont {Hart}, \citenamefont {Hitchcock}, \citenamefont {Corcovilos}, \citenamefont {Yang}, \citenamefont {Reed},\ and\ \citenamefont {Hulet}}]{Duarte2011All}%
  \BibitemOpen
  \bibfield  {author} {\bibinfo {author} {\bibfnamefont {P.~M.}\ \bibnamefont {Duarte}}, \bibinfo {author} {\bibfnamefont {R.~A.}\ \bibnamefont {Hart}}, \bibinfo {author} {\bibfnamefont {J.~M.}\ \bibnamefont {Hitchcock}}, \bibinfo {author} {\bibfnamefont {T.~A.}\ \bibnamefont {Corcovilos}}, \bibinfo {author} {\bibfnamefont {T.-L.}\ \bibnamefont {Yang}}, \bibinfo {author} {\bibfnamefont {A.}~\bibnamefont {Reed}},\ and\ \bibinfo {author} {\bibfnamefont {R.~G.}\ \bibnamefont {Hulet}},\ }\href {https://doi.org/10.1103/PhysRevA.84.061406} {\bibfield  {journal} {\bibinfo  {journal} {Phys. Rev. A}\ }\textbf {\bibinfo {volume} {84}},\ \bibinfo {pages} {061406} (\bibinfo {year} {2011})}\BibitemShut {NoStop}%
\bibitem [{\citenamefont {Grier}\ \emph {et~al.}(2013)\citenamefont {Grier}, \citenamefont {Ferrier-Barbut}, \citenamefont {Rem}, \citenamefont {Delehaye}, \citenamefont {Khaykovich}, \citenamefont {Chevy},\ and\ \citenamefont {Salomon}}]{grier2013lambda}%
  \BibitemOpen
  \bibfield  {author} {\bibinfo {author} {\bibfnamefont {A.~T.}\ \bibnamefont {Grier}}, \bibinfo {author} {\bibfnamefont {I.}~\bibnamefont {Ferrier-Barbut}}, \bibinfo {author} {\bibfnamefont {B.~S.}\ \bibnamefont {Rem}}, \bibinfo {author} {\bibfnamefont {M.}~\bibnamefont {Delehaye}}, \bibinfo {author} {\bibfnamefont {L.}~\bibnamefont {Khaykovich}}, \bibinfo {author} {\bibfnamefont {F.}~\bibnamefont {Chevy}},\ and\ \bibinfo {author} {\bibfnamefont {C.}~\bibnamefont {Salomon}},\ }\href {https://doi.org/10.1103/PhysRevA.87.063411} {\bibfield  {journal} {\bibinfo  {journal} {Phys. Rev. A}\ }\textbf {\bibinfo {volume} {87}},\ \bibinfo {pages} {063411} (\bibinfo {year} {2013})}\BibitemShut {NoStop}%
\bibitem [{\citenamefont {Salomon}\ \emph {et~al.}(2014)\citenamefont {Salomon}, \citenamefont {Fouch{\'e}}, \citenamefont {Wang}, \citenamefont {Aspect}, \citenamefont {Bouyer},\ and\ \citenamefont {Bourdel}}]{salomon2014gray}%
  \BibitemOpen
  \bibfield  {author} {\bibinfo {author} {\bibfnamefont {G.}~\bibnamefont {Salomon}}, \bibinfo {author} {\bibfnamefont {L.}~\bibnamefont {Fouch{\'e}}}, \bibinfo {author} {\bibfnamefont {P.}~\bibnamefont {Wang}}, \bibinfo {author} {\bibfnamefont {A.}~\bibnamefont {Aspect}}, \bibinfo {author} {\bibfnamefont {P.}~\bibnamefont {Bouyer}},\ and\ \bibinfo {author} {\bibfnamefont {T.}~\bibnamefont {Bourdel}},\ }\href {https://doi.org/10.1209/0295-5075/104/63002} {\bibfield  {journal} {\bibinfo  {journal} {Europhys.\ Lett.}\ }\textbf {\bibinfo {volume} {104}},\ \bibinfo {pages} {63002} (\bibinfo {year} {2014})}\BibitemShut {NoStop}%
\bibitem [{\citenamefont {Pucher}(2018)}]{pucher2018spectroscopy}%
  \BibitemOpen
  \bibfield  {author} {\bibinfo {author} {\bibfnamefont {S.}~\bibnamefont {Pucher}},\ }\emph {\bibinfo {title} {Spectroscopy of the 6S$_{1/2}\rightarrow$5D$_{5/2}$ Electric Quadrupole Transition of Atomic Cesium}},\ \href {https://www.physik.hu-berlin.de/de/gop/publications/2018_10_27_master_thesis_pucher_compressed.pdf} {\bibinfo {type} {Diplomarbeit}},\ \bibinfo  {school} {Technische Universit{\"a}t Wien}, \bibinfo {address} {Vienna, Austria} (\bibinfo {year} {2018}),\ \bibinfo {note} {advisor: Univ. Prof. Dr. Arno Rauschenbeutel}\BibitemShut {NoStop}%
\bibitem [{\citenamefont {Woolley}(2020)}]{wooley2020power}%
  \BibitemOpen
  \bibfield  {author} {\bibinfo {author} {\bibfnamefont {R.~G.}\ \bibnamefont {Woolley}},\ }\href {https://doi.org/10.1103/PhysRevResearch.2.013206} {\bibfield  {journal} {\bibinfo  {journal} {Phys. Rev. Res.}\ }\textbf {\bibinfo {volume} {2}},\ \bibinfo {pages} {013206} (\bibinfo {year} {2020})}\BibitemShut {NoStop}%
\end{thebibliography}%

\end{document}